\documentclass[12pt]{iopart2}

\usepackage{epsfig,amssymb,amsmath}

\def\spose#1{\hbox to 0pt{#1\hss}}
\def\lesssim{\mathrel{\spose{\lower 3pt\hbox{$\mathchar"218$}}
 \raise 2.0pt\hbox{$\mathchar"13C$}}}
\def\gtrsim{\mathrel{\spose{\lower 3pt\hbox{$\mathchar"218$}}
 \raise 2.0pt\hbox{$\mathchar"13E$}}}

\def\<{\langle}
\def\>{\rangle}

\begin{document}

\title{ 
The quantum XY chain with boundary fields: Finite-size gap and phase behavior
}

\author{Aldo Coraggio$^1$ and Andrea Pelissetto$^2$}
\address{$^1$ SISSA, Via Bonomea 265, I-34136 Trieste, Italy \\
         $^2$ Dipartimento di Fisica di Sapienza, Universit\`a di Roma
        and INFN, Sezione di Roma I, Piazzale A. Moro 2, I-00185 Roma, Italy}

\ead{
acoraggi@sissa.it,
Andrea.Pelissetto@roma1.infn.it}

\begin{abstract}
We present a detailed study of the finite-size one-dimensional 
quantum XY chain in a transverse field 
in the presence of boundary fields coupled with the
order-parameter spin operator.
We consider fields located at the chain boundaries
that have the same strength and that are  oppositely aligned.
We derive exact expressions for the gap $\Delta$ 
as a function of the model parameters
for large values of the chain length $L$.
These results allow us to characterize the nature of the 
ordered phases of the model. We find a magnetic (M) phase 
($\Delta \sim e^{-aL}$), a magnetic-incommensurate (MI) 
phase ($\Delta \sim e^{-aL} f_{MI}(L)$),  a kink (K) 
phase ($\Delta \sim L^{-2}$), and a 
kink-incommensurate (KI) phase ($\Delta \sim L^{-2} f_{KI}(L)$);
$f_{MI}(L)$ and $f_{KI}(L)$ are bounded oscillating functions of $L$. 
We also analyze the behavior along the phase boundaries. In particular,
we characterize the universal crossover behavior across the K-KI phase
boundary. On this boundary, the dynamic
critical exponent is $z=4$, i.e., $\Delta \sim L^{-4}$ for large values of
$L$.
\end{abstract}


\maketitle


\section{Introduction}
\label{intro}

Understanding how classical or quantum many-body systems order under 
the action of an external field is an important problem in condensed-matter
physics. These ordering phenomena are signalled by phase transitions, which,
both in a classical or quantum setting, can be broadly classified in two 
categories. First, there are continuous transitions characterized 
by long-range correlations decaying generically as powers of the 
distance and by 
a universal large-scale behavior. In this case, many features---for
instance, the nature of the two phases that are separated by the 
transition---are 
independent of the microscopic details and can be determined in 
simplified models that are only characterized by a few properties,
like the global and local symmetries, the dimensionality of the 
order parameter, and the nature of the symmetry breaking pattern 
at the transition. 
In a  finite-size system, long-distance 
global quantities have a nonanalytic dependence on the system size, which is 
also independent of the specific nature of the 
interactions and of the boundary conditions, which,
however, may affect scaling functions and universal amplitudes,
see, e.g., Ref.~\cite{CPV-14}.
A second class of transitions are the first-order ones, characterized 
by the discontinuity of thermodynamic and global observables. From the point
of view of phase behavior, discontinuous transitions are more interesting, as 
their phase behavior is more diverse. For instance, 
the nature of the coexisting phases crucially depends on the 
nature of the boundary conditions, even in the infinite-volume limit,
see, e.g., Ref.~\cite{PV-24}.
Therefore, by simply varying the 
boundary interactions one can generate a variety of different bulk behaviors.

In this work we consider the one-dimensional quantum XY model in 
a transverse magnetic field 
\cite{LSM-61,Katsura-62,Niemeijer-67,Niemeijer-68,Pfeuty-70,BMD-70,%
BM-71,Suzuki-71}, a paradigmatic integrable system, for which it is possible
to obtain exact results for many ground-state properties, using its
relation with a model of nonlocal free fermions.
The Hamiltonian of an XY chain of length $L$ is given by 
\begin{eqnarray}
H  = 
- {1\over 2} \sum_{i=1}^{L-1} 
     [ (1 + \gamma)\sigma^{(1)}_i \sigma^{(1)}_{i+1} +
       (1 - \gamma) \sigma^{(2)}_i \sigma^{(2)}_{i+1} ]
- g \sum_{i=1}^L \sigma^{(3)}_i ,
\label{Isc}
\end{eqnarray}
where $\sigma^{(i)}$ are the Pauli matrices. It is easy to verify that 
the spectrum is invariant under $g\to -g$, so we can set $g\ge 0$
without loss of generality. For periodic or open boundary conditions,
the model  is also invariant under $\gamma \to -\gamma$. 
However, this symmetry is broken by the boundary conditions we will use
and therefore we will consider positive and negative values of $\gamma$. 
For $\gamma = \pm 1$, we obtain the simpler Ising chain, while 
for $\gamma = 0$, the Hamiltonian becomes that of the XX chain with 
an enlarged U(1) symmetry.

For $g = 1$ and any $\gamma$ the system undergoes a 
continuous quantum transition
that separates a paramagnetic (disordered) phase ($g > 1$) from 
an ordered phase with degenerate ground state ($g < 1$). 
The nature of the latter phase depends on 
the boundary conditions. For periodic and open boundary conditions
the phase diagram for $g < 1$ is well known,
see, e.g., Ref.~\cite{Franchini-17}.  For $g^2 + \gamma^2 > 1$ 
there is an ordinary ferromagnetic phase: In the infinite-volume limit
the ground state is doubly degenerate. This degeneracy is lifted 
in a finite volume, with a gap that behaves as $e^{-a L}$.
On the other hand, for $g^2 + \gamma^2 < 1$ an oscillatory phase appears:
the energy gap 
shows an oscillating behavior as a function of the system size,
and correlation functions show oscillations as a function of  the 
distance.  The behavior for $g^2 + \gamma^2 = 1$ is somewhat peculiar, 
as the ground-state wavefunction factorizes into a product of single spin 
states \cite{KTM-82,MS-85}.

If boundary fields are added, the phase behavior for $g < 1$ 
becomes more complex. 
The analysis of Refs.~\cite{CPV-15,CPV-15JSTAT,Hu2021} 
for the Ising chain ($\gamma=1$)
shows that the addition of 
oppositely oriented longitudinal magnetic fields at the chain 
boundaries stabilizes a new phase,
named kink phase. In this phase there is no ferromagnetic order and the 
low-energy excitations are propagating kink states \cite{Sachdev-book} 
of momentum of order $1/L$.  The kink phase and the ferromagnetic phase 
are separated by a continuous transition, with a universal crossover 
behavior. Using the quantum-to-classical mapping 
one can relate this transition to the wetting transition
\cite{Dietrich-88,Indekeu-94,BR-01,BLM-03} that occurs in
classical two-dimensional Ising systems in a strip geometry 
\cite{Abraham-80,NF-82,Ciach-86,CS-87,PS-88,PE-90,PEN-91,SMO-94,MS-96,
Maciolek-96,AM-10}.

The finite-size behavior of the Ising and of the XY chain has 
been extensively studied 
\cite{BS-74,CJ-87,BC-87,Henkel-87,BG-87,BT-90,
TB-93,IPT-93,Karewski-00,
IH-01,Peschel-04,IH-09,DDSCRA-10,CPV-14,CPV-15,
CPV-15JSTAT,Hu2021,Hu2023}.
In this work we study the finite-size behavior of the gap in the 
presence of oppositely oriented boundary longitudinal fields (OBF), 
extending the results obtained in 
Refs.~\cite{CPV-15JSTAT,Hu2021} for the Ising chain to the XY model.
By combining analytic and numerical methods, we 
obtain exact results for the large-size behavior of the low-energy
excitations of the model, which, 
in turn, allow us to determine the different possible phases for $g < 1$.

The two phases that occur when periodic boundary conditions are used,
the conventional ferromagnetic phase (we name it magnetized (M) phase) 
and the oscillatory phase (named magnetized-incommensurate (MI) phase)
also occur in the presence of OBF. In both cases the gap decreases
exponentially, with additional size oscillations in the MI phase. 
If the boundary fields are sufficiently strong, we find a kink (K) 
phase, as in the Ising chain \cite{CPV-15JSTAT}, with delocalized excitations
and a gap that decreases as $1/L^2$. Finally, we find a novel phase, that 
we name kink-incommensurate (KI) phase, in which excitations are 
delocalized, so the gap decreases as $1/L^2$, but which is also characterized 
by incommensurate oscillations, as the MI phase. We have also 
studied the behavior of the gap along the boundaries that separate the 
different phases. In particular, along the K-KI boundary we find that the 
gap scales as $L^{-4}$, i.e., with a dynamical critical exponent $z=4$. 

Finally, we discuss  the crossover behavior that is observed when parameters 
are varied across a phase boundary. We have considered the crossover 
between the M and the K phase, obtaining the same behavior as 
observed in the Ising chain. As expected, the crossover 
across the M-K boundary is universal. We also discuss the behavior 
across the K-KI boundary. Also in this case we find a universal scaling regime. 
We determine the appropriate scaling variable and compute 
the scaling function for the energy gap.

The paper is organized as follows.  In Sec.~\ref{sec2} we introduce
the one-dimensional quantum XY chain with boundary fields. 
In Secs.~\ref{sec3} and \ref{sec4}
we compute the low-energy spectrum by generalizing the approach 
of Ref.~\cite{Hu2021} and exploiting the equivalent
quadratic fermionic formulation of the Hamiltonian 
\cite{LSM-61,Pfeuty-70}.  The different phases are discussed
in the subsequent sections. In Sec.~\ref{sec5} we sketch 
the phase diagram as a function of the model parameters and 
characterize the behavior of the low-energy excitations in the 
different phases. In Sec.~\ref{sec6} we discuss the magnetized phases
and in Sec.~\ref{sec7} the kink phase. In Sections~\ref{sec8} and 
\ref{sec9} we discuss the crossover behavior across the M-K and K-KI 
boundaries.  In Sec.~\ref{sec10} we present our conclusions.
Technical details are reported in the Appendices.

\section{Model and definitions}
\label{sec2}

In this work we focus on the low-energy spectrum of the XY model with 
Hamiltonian ~(\ref{Isc}).
It is important to observe that the model is 
ferromagnetic for any value of $\gamma$. 
For $|\gamma| \le 1$, both hopping terms favor the alignment of 
the neighboring spins. For $|\gamma| > 1$, instead, the two terms have opposite 
signs, i.e., one is ferromagnetic and one is antiferromagnetic. The 
ferromagnetic interaction, however, is always the dominant one.

The XY chain undergoes a continuous transition at $g=1$ 
\cite{LSM-61,Katsura-62},
separating a quantum ordered phase ($g<1$) from a quantum paramagnetic 
phase ($g>1$). In this paper 
we investigate the effects of boundary magnetic fields aligned along
the $x$ axis. They give rise to an additional energy term
\begin{equation}
H_b = - \zeta_1 \sigma_1^{(1)} - \zeta_L \sigma_L^{(1)},
\label{hb}
\end{equation} 
which is added to Hamiltonian (\ref{Isc}). In the following we  
only consider the case of oppositely-aligned boundary fields (OBF), that
correspond to 
\begin{equation}
\zeta_1 = -\zeta_L. 
\label{obfdef}
\end{equation}
We focus on the low-energy spectrum of the model. 
In particular, we will obtain exact finite-size results for the 
energy differences between the lowest states and the ground state
\begin{equation}
\Delta_{n}\equiv E_n-E_0, 
\label{deltaldef}
\end{equation}
(here 
$E_n$ are the energy eigenvalues ordered so that $E_0\le E_1\le E_2\ldots$)
and, in particular, for the finite-size gap $\Delta = E_1 - E_0$.

\section{Jordan-Wigner representation and Hamiltonian diagonalization}
\label{sec3}

To determine the spectrum of Hamiltonian (\ref{Isc}), we use the technique
introduced in Ref.~\cite{CPV-15JSTAT}.  We extend the
model, considering two additional spins located in 0 and $L+1$ and the
Hamiltonian
\begin{equation}
H_e = - {1\over 2} \sum_{i=1}^{L-1} 
    [(1 + \gamma)\sigma_i^{(1)} \sigma^{(1)}_{i+1} + 
     (1 - \gamma)\sigma_i^{(2)} \sigma^{(2)}_{i+1}] - 
    J_0 \,\sigma_0^{(1)} \sigma_1^{(1)} - 
    J_L \,\sigma_{L}^{(1)} \sigma_{L+1}^{(1)}  - 
    g \sum_{i=1}^{L} \sigma_i^{(3)}.
\label{def-Hextended}
\end{equation}
This is the XY Hamiltonian with two different couplings on the
boundary links and zero transverse field on the boundaries.  
To compute the spectrum of Hamiltonian (\ref{def-Hextended}), 
we follow Ref.~\cite{CPV-15JSTAT}. We first
perform a Jordan-Wigner transformation and 
then we perform a Bogoliubov transformation (see 
\ref{App.A} for details). The Hamiltonian takes the form
\begin{equation}
   H_e = E_{gs} + \sum_{k=0}^{L+1} {\cal E}_k \eta^\dagger_k \eta_k,
\label{H1-diag}
\end{equation}
with $0\le {\cal E}_0 \le {\cal E}_1 \le \ldots$, where $\eta_k$ 
are canonical fermionic operators, and 
\begin{equation}
   E_{gs} = - {1\over2} \sum_{k=0}^{L+1} {\cal E}_k.
\end{equation}
The squared energies ${\cal E}_k^2$ of the 
fermionic modes are the eigenvalues of the matrix $C$ given by
\begin{equation}
C=
 \begin{pmatrix}
    e & h & d \\
    h & f & c & b \\
    d & c & a & c & b \\
    0 & b & c & a & c & b  \\
      &   & \ddots & \ddots & \ddots & \ddots & \ddots & & \vdots \\ 
      & &        & b & c & a & c & b & 0\\
      & &        &   & b & c & a & c & 0 \\
      & &        &   &   & b & c & l & 0 \\
      & &        &   & \dots   & 0 & 0 &0&0
  \label{eq:defCmatrix}
  \end{pmatrix}
\end{equation}
with
\begin{eqnarray}
   a &=& 2(1+\gamma^2)+4g^2 \nonumber \\
   b &=& 1-\gamma^2         \nonumber \\
   c &=& 4g                 \nonumber \\
   d &=& 2J_0(1-\gamma)     \nonumber \\
   e &=& 4J_0^2             \nonumber \\
   f &=& 4g^2+(1+\gamma)^2  \nonumber \\
   h &=& 4J_0 g             \nonumber \\
   l &=&4J_L^2+4g^2+(1-\gamma)^2\,. 
\end{eqnarray}
The matrix $C$ has a zero eigenvalue, ${\cal E}_0^2 = 0$, that is 
related to the double degeneracy of the spectrum of $H_e$ (see
\ref{App.A} for a more detailed discussion).
To determine the non-zero eigenvalues, we consider the square 
matrix $\hat{C}$ of size $(L+1)\times (L+1)$, that is 
obtained from $C$ by deleting the last column and the last row. 
The gap in the presence of OBF is  expressed \cite{CPV-15JSTAT} in 
terms of the two lowest eigenvalues of $\hat{C}$, ${\cal E}_1^2$ and
${\cal E}_2^2$, computed setting  $J_0 = J_L = \zeta_1= - \zeta_L$, as   
\begin{equation}
  \Delta_{OBF} = {\cal E}_2 - {\cal E}_1,
\end{equation}

\section{Determination of the spectrum} \label{sec4}

To determine the spectrum of the matrix $\hat{C}$, we extend 
the results of Ref.~\cite{Hu2021,Hu2023}. Ref.~\cite{Hu2021}
introduced a parametrization
of the eigenvectors of the matrix $\hat{C}$ that is exact for the 
Ising chain ($\gamma = 1$). This parametrization was generalized in 
Ref.~\cite{Hu2023} to discuss the spectrum of the XY chain. 
Although it does not provide exact eigenvectors 
of $\hat{C}$ for finite values of $L$,
Ref.~\cite{Hu2023} argued that the approximation
becomes exact in the infinite-size limit, if 
the eigenvectors are localized (if the ground state is magnetized in the 
language we use in this paper). Here we generalize the approach of 
Ref.~\cite{Hu2021}, obtaining the exact eigenvectors $\psi_L$
of $\hat{C}$ for any finite $L$. By definition, they satisfy the 
eigenvalue equations 
\begin{equation}
  \hbox{Eq}_k = \sum_i \hat{C}_{ki} \psi_{L,i} - {\cal E}^2 \psi_{L,k} = 0,
\label{Eqk}
\end{equation}
with $k,i = 1,\ldots, L+1$. 

To determine $\psi_L$, we first note that a vector 
of components 
\begin{eqnarray}
\overline{\psi}_{L,1} = \frac{1+\gamma}{2J_0} A (-x) \qquad
\overline{\psi}_{L,k} = A (-x)^k \quad \hbox{for $k\geq 2$},
\label{psi-simple}
\end{eqnarray}
($x$ and $A$ are arbitrary complex numbers)
satisfies the equations (\ref{Eqk}) for any $k = 3,\ldots,L-1$, provided we 
identify ${\cal E}^2 = \varepsilon^2(x)$, where
\begin{equation}
    \varepsilon^2(x)=(1-\gamma^2)(x+x^{-1})^2-4g(x+x^{-1})+4(g^2+\gamma^2)\,.
    \label{eq:defepssq}
\end{equation}
The equations (\ref{Eqk}) for the boundary components $k=1,2,L,L+1$ 
instead are not satisfied
and therefore this vector is not an eigenvector of $\hat{C}$. 
To obtain an eigenvector we define $\psi_L$ as a
linear combination of vectors of the form reported above.
The vector $\psi_L$ 
depends on a few parameters that are fixed by requiring Eq.~(\ref{Eqk}) to 
also hold on the boundaries, i.e., for $k=1,2,L,L+1$. 

Specifically, as in Ref.~\cite{Hu2023}, we consider two 
different complex numbers $x_1$ and $x_2$. Then, we parametrize the 
eigenvectors of $\hat{C}$ as \footnote{To make contact with 
Eq.~(\ref{psi-simple}), note that the terms with coefficient $c_1$ 
correspond to a vector $\bar{\psi}_L$ with $A = - c_1 x_1^3$ and $x= 1/x_1$,
while those with coefficient $d_1$ correspond to a vector $\bar{\psi}_L$ with 
$A = - d_1 x_1^{2-L}$ and $x = x_1$.}
\begin{equation}
\begin{aligned}
     \psi_{L,1} &=  \frac{1+\gamma}{2J_0} 
     (c_1 x_1^2+c_2 x_2^2+d_1 x_1^{3-L}+d_2 x_2^{3-L})\,, \\
     \psi_{L, i} &=  (-1)^{i-1}
     (c_1 x_1^{3-i}+c_2 x_2^{3-i}+d_1 x_1^{2-L+i}+d_2 x_2^{2-L+i})\,, \\
\end{aligned}
\label{psi-Ansatz}
\end{equation}
where $i=2,\ldots,L+1$. 
With the parametrization  (\ref{psi-Ansatz}), 
equations $\hbox{Eq}_k=0$ with 
$k = 3,\ldots,L-1$ are exactly satisfied for any $c_1$, $c_2$, $d_1$ and $d_2$,
provided that 
${\cal E}^2 = \varepsilon^2(x_1) = \varepsilon^2(x_2)$.
If we require $\psi_L$ to be an eigenvector of $\hat{C}$, also 
equations $\hbox{Eq}_k=0$ for $k=1,2,L,L+1$ should be satisfied. 
Explicitly they can be written as:
\begin{eqnarray}
\hbox{Eq}_1 &= & c_1 E(x_1) F(x_1)+c_2 E(x_2) F(x_2) + \nonumber \\
   && \qquad 
   d_1 x_1^{5-L} E(x_1^{-1}) F(x_1^{-1})+
   d_2 x_2^{5-L} E(x_2^{-1}) F(x_2^{-1}) =0, \nonumber \\
\hbox{Eq}_2 &= & c_1 E(x_1) +c_2 E(x_2) + \nonumber \\
&& \qquad d_1 x_1^{5-L}E(x_1^{-1}) +d_2 x_2^{5-L} E(x_2^{-1}) =0, 
\label{eq:foureqs} \\
\hbox{Eq}_{L} &=& (1-\gamma^2)
   (c_1 x_1^{1-L}+c_2 x_2^{1-L}+d_1 x_1^{4}+d_2 x_2^{4} )= 0 , \nonumber \\
\hbox{Eq}_{L+1} &=&  d_1 A(x_1)+d_2 A(x_2)+
    c_1 A(x_1^{-1}) x_1^{3-L}+c_2 A(x_2^{-1}) x_2^{3-L} = 0 ,
\nonumber 
\end{eqnarray}
with 
\begin{eqnarray}
A(x)&=&x^2 [(1+\gamma)^2+x^2 (1-\gamma^2)-4 g x-4 J_L^2]\, , \nonumber \\
E(x)&=& x^3 (1-\gamma)[1+\gamma-2g x^{-1}+(1-\gamma)x^{-2}]\,, \\
F(x)&=& -\frac{(\gamma +1) \left(-\gamma +2 g x+(\gamma -1)
x^2-1\right)+4 J_0^2}{2 (\gamma -1) J_0 x} \,. \nonumber 
\end{eqnarray}
Let us note that the second and the third equation
in Eq.~(\ref{eq:foureqs}) are trivial for $\gamma = 1$, 
since in the transverse-field Ising chain there are only two 
nontrivial equations, see Ref.~\cite{Hu2021}.

Equations~(\ref{eq:foureqs}) together with the constraint
$\varepsilon^2(x_1)=\varepsilon^2(x_2)$ allow us to determine the 
parameters $x_1$ and $x_2$ and the coefficients $c_1$, $c_2$, $d_1$, 
and $d_2$ (up to a common multiplicative constant) that make $\psi_L$
an eigenvector of $\hat{C}$.\footnote{This procedure allows us to obtain
vectors $\psi_L$ that are 
eigenvectors of $\hat{C}$ for finite $L$. 
In \ref{App.consistency} we prove that 
this approach provides all eigenvectors of $\hat{C}$.} Once these quantities 
are determined, the energies are obtained using 
${\cal E}^2 = \varepsilon(x_1)^2$ or, equivalently, 
${\cal E}^2 = \varepsilon(x_2)^2$.

Note that the parametrization is invariant under the exchange of 
$x_1$ and $x_2$ and under a second set of symmetries. If we
define
\begin{equation}
   c'_1 = d_1 x_1^{5-L} \qquad d'_1 = c_1 x_1^{5-L},
\end{equation}
we have
\begin{equation}
c_1 x_1^{3-i}+d_1 x_1^{2-L+i} =
  c'_1 \left({1\over x_1}\right)^{3-i} +
  d'_1 \left({1\over x_1}\right)^{2-L+i},
\end{equation}
which shows that the parametrization is invariant under $x_1 \to 1/x_1$.
The same holds for the parameter $x_2$.
The presence of these symmetries allows us to 
always choose $x_1$ and $x_2$
such that $|x_1| \ge 1$, $|x_2| \ge 1$, and $|x_1|\ge |x_2|$.
Moreover, $\psi_{L}$ is
defined up to a multiplicative constant. Thus, we can arbitrarily set one
of the parameters $c_1$, $c_2$, $d_1$, and $d_2$ equal to one or zero.

It is important to stress that the relevant parameters for the 
analysis of the spectrum are the two complex numbers $x_1$ and $x_2$.
Indeed, each eigenvalue ${\cal E}^2$ is uniquely related with 
a pair $x_1$, $x_2$ of solutions of the equation 
${\cal E}^2 = \varepsilon(x)^2$.\footnote{In some cases we will
need to explicitly specify to which level the values $x_1$ and $x_2$
correspond. As we label the eigenvalues of $\hat{C}$
with a positive integer $n$ (the $n$-th eigenvalue is 
${\cal E}_n^2$), we refer to the corresponding values of $x_1$ and $x_2$
as $x_{1,n}$ and $x_{2,n}$. An analogous notation
will be used for the coefficients that parametrize the large-$L$
behavior of $x_1$ and $x_2$. Thus, if we define 
$x_2 \approx \exp (i \theta_1/L)$, $\theta_{1,n}$ indicates the value 
of $\theta_1$ for the $n$-th level of $\hat{C}$.}
Since the parametrization is invariant under 
$x_i\to 1/x_i$ and under the exchange of $x_1$ and $x_2$, 
$x_1$ and $x_2$ can be identified with 
any pair of solutions that are not related by the inversion 
symmetry. 

To make contact with the solution in the presence of periodic boundary 
conditions, see, e.g., Ref.~\cite{Franchini-17}, note that, if $x = e^{iq}$,
the function $\varepsilon(x)$ can be rewritten as 
\begin{equation}
  \varepsilon(e^{iq})^2 = 4 [( g - \cos q)^2 + \gamma^2 \sin^2 q],
\label{dispersion}
\end{equation}
which is the dispersion relation between the momentum $q$ and the 
energy for the XY model with periodic boundary
conditions. Thus, in our parametrization, the eigenstates are 
linear combinations of two different excitations, whose momentum 
is encoded in the parameters $x_1$ and $x_2$. These excitations
are delocalized if $|x_i|= 1$, localized in the opposite case.

In the following we will study the phase diagram of the model for 
$J_0>0$. The case $J_0 = 0$ (open boundary conditions) 
can also be addressed with the same method, but requires a priori a separate 
treatment, since our parametrization becomes singular in the limit 
$J_0\to 0$, see Eq.~(\ref{psi-Ansatz}). In practice, only the coefficients
$c_i$ and $d_i$ are singular for $J_0 \to 0$. The parameters 
$x_i$ are continuous and thus all results for the phase diagram 
also apply for $J_0 = 0$, i.e., for open boundary conditions. 
We only consider the case $0 < g < 1$,
in which the bulk behavior depends on the boundary conditions.
Indeed, for $g > 1$ the system is paramagnetic and the gap is finite, while,
for $g = 1$, the behavior is independent of the microscopic details and 
thus the value of $\gamma$ only affects nonuniversal constants and 
scaling corrections \cite{CPV-14}.

\section{Numerical determination of the phase diagram}
\label{sec5}

\begin{figure}[tbp]
\begin{center}
\begin{tabular}{cc}
\epsfig{width=8truecm,angle=0,file=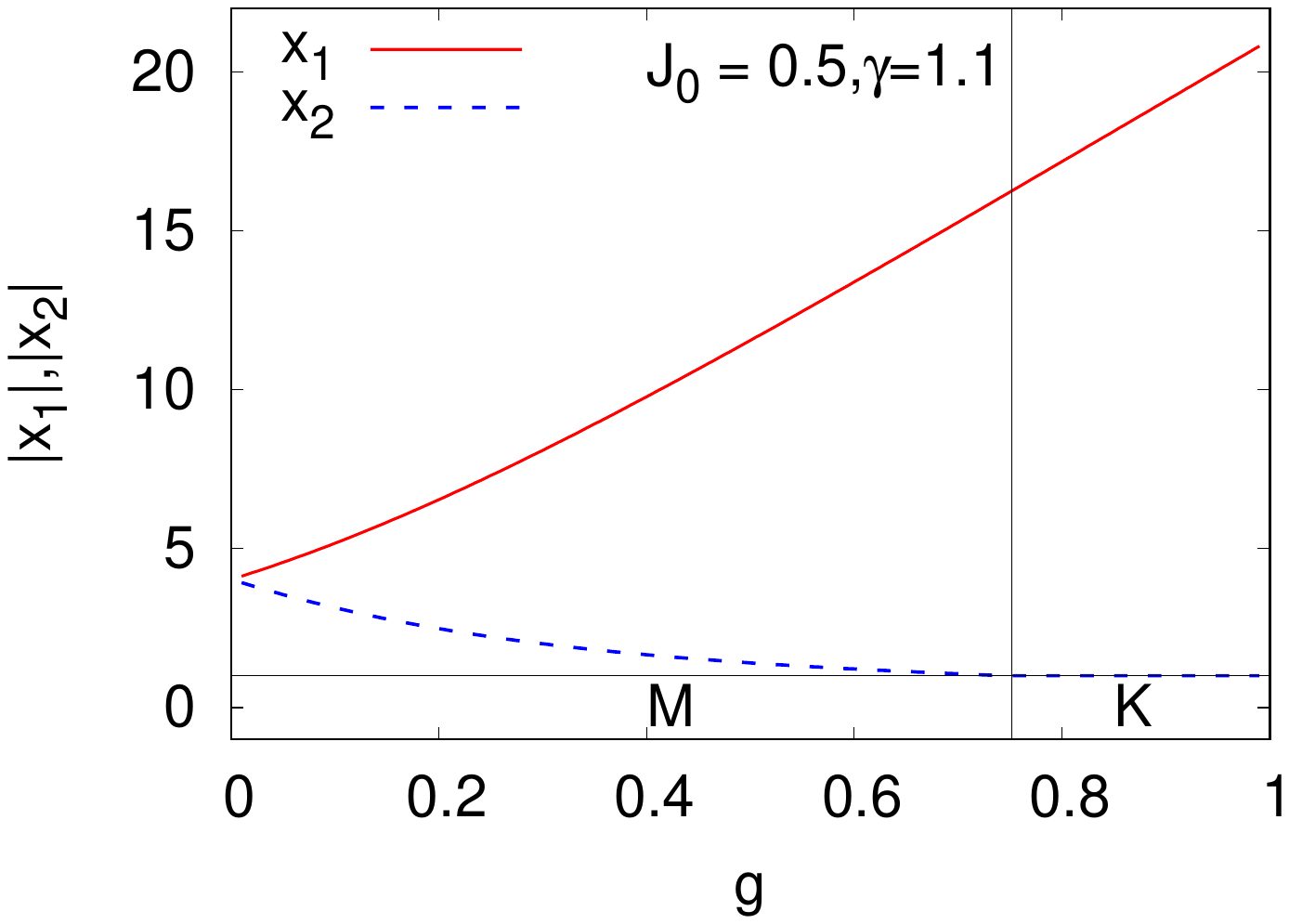} & \hspace{-1cm}
\epsfig{width=8truecm,angle=0,file=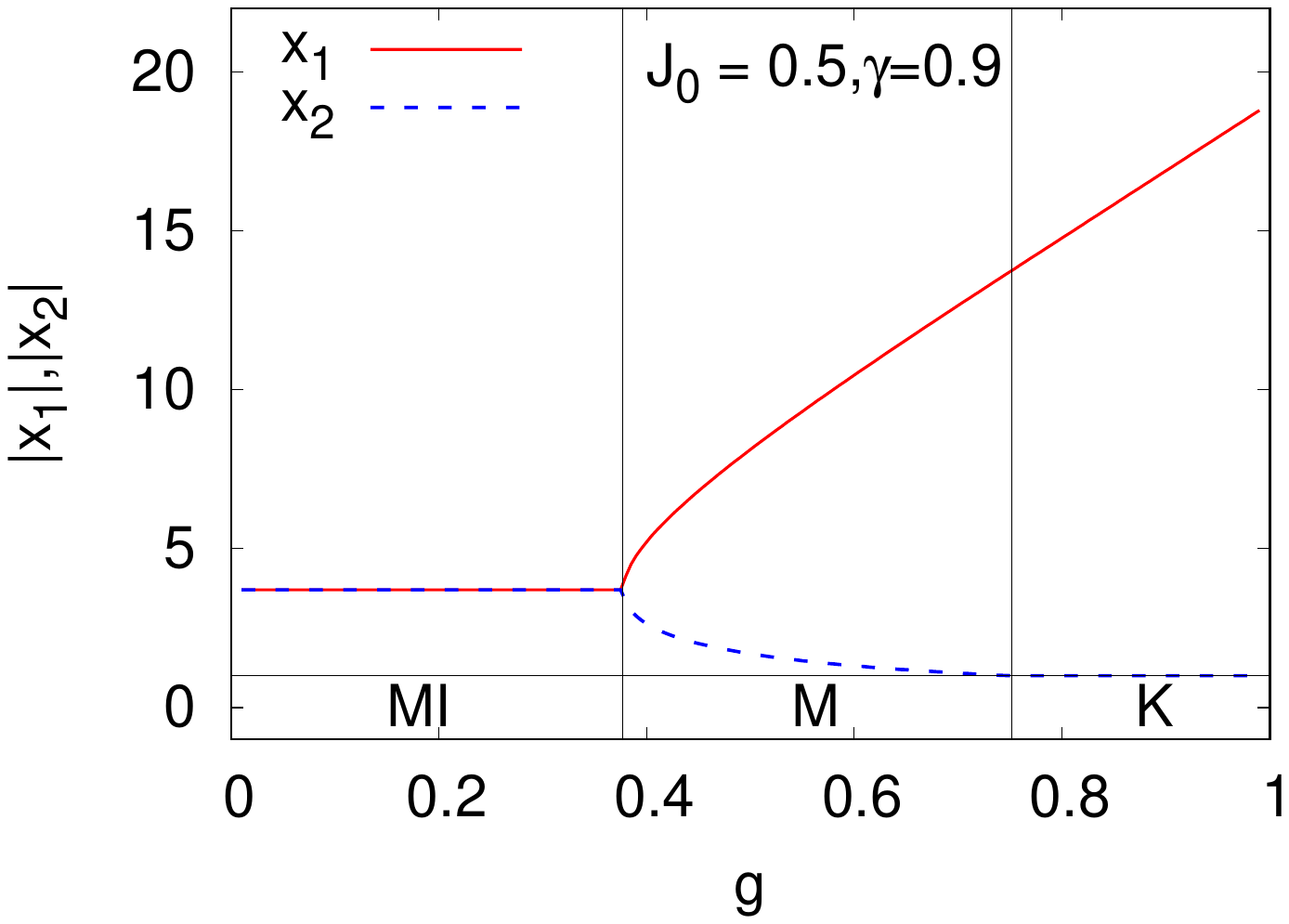} \\
\epsfig{width=8truecm,angle=0,file=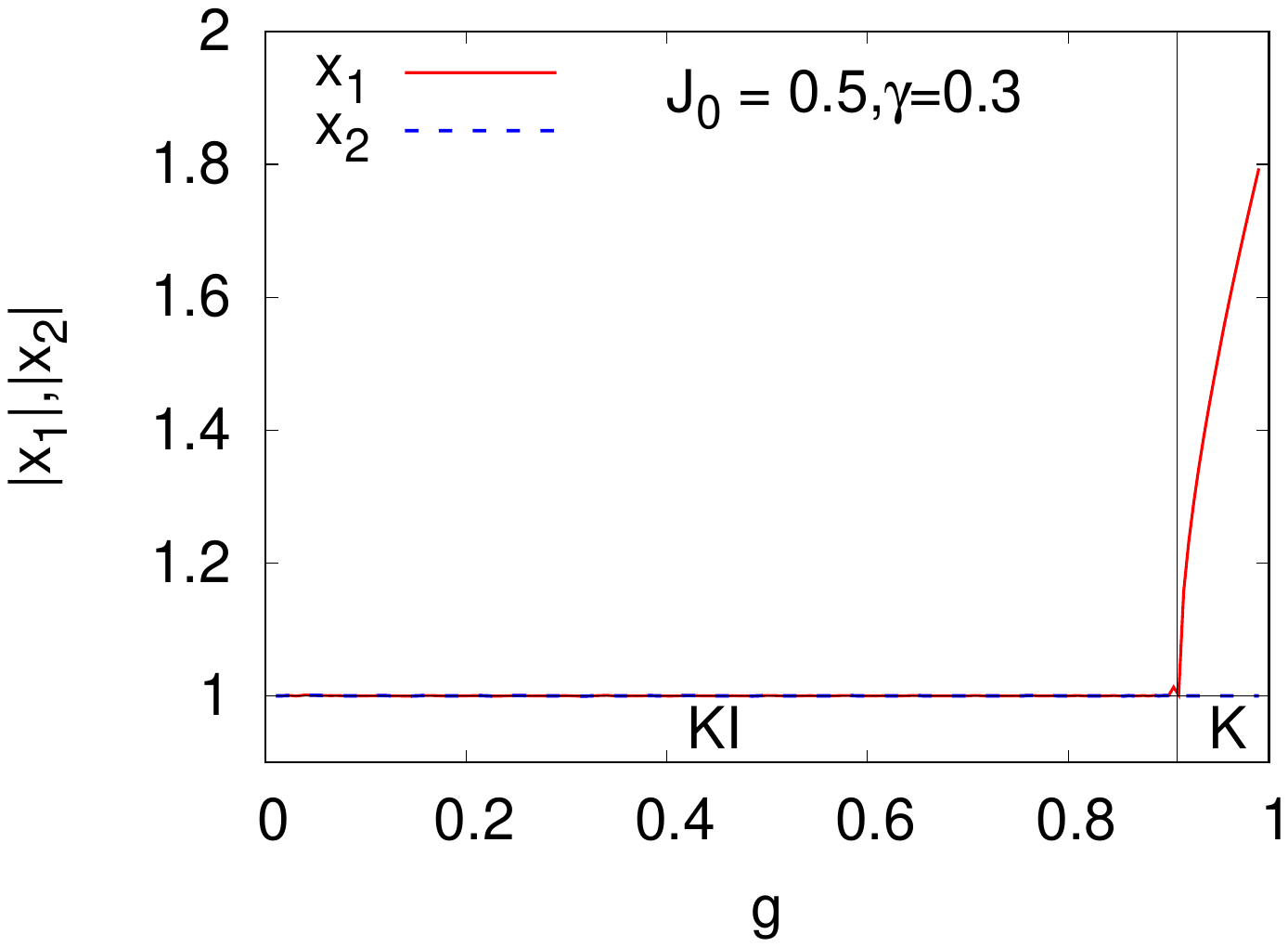} & \hspace{-1cm}
\epsfig{width=8truecm,angle=0,file=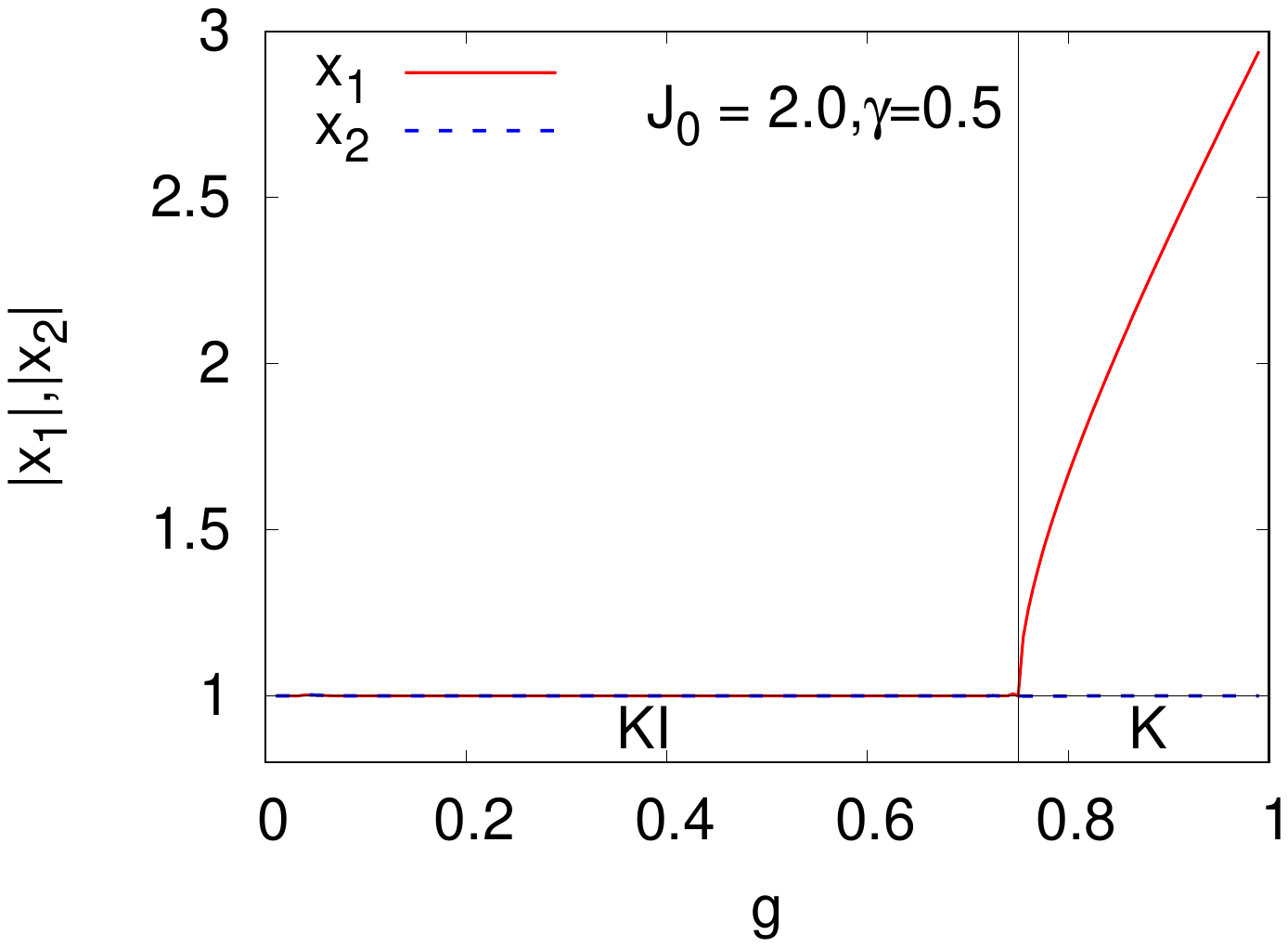} \\
\end{tabular}
\end{center}
\vskip-5mm
\caption{Plots of $|x_1|$ and $|x_2|$ as a function of 
$g$, for given $J_0$ and $\gamma$. a) (Top left) $x_1$ and $x_2$ are 
both real (M phase, type (i) solution) for $g\lesssim 0.755$, 
while $x_1$ is real and 
$x_2$ is complex (K phase, type (iii) solution) in the opposite case.
b) (Top right) $x_1$ and $x_2$ are both complex 
(MI phase, type (ii) solution) with 
$|x_1| = |x_2| > 1$ up to $g \approx 0.377$; then, they become real
(M phase, type (i) solution) up to $g \approx 0.752$; 
for larger values of $g$, $x_1$ is real and $x_2$ is complex 
(K phase, type (iii) solutions).
c) (Bottom) In both cases for small $g$, we have $|x_1|=|x_2| = 1$
(KI phase, type (iv) solution); 
then $x_1$ becomes real, while $x_2$ is complex (K phase, type (iii) solution);
the transition occurs for  $g\approx 0.908$ (left) and $g = 0.75$
(right).
}
\label{x1x2-genericJ0gamma}
\end{figure}

As we have already discussed, the parameters $x_1$ and $x_2$ satisfy 
${\varepsilon}(x)^2 = {\cal E}^2$. Let us first discuss some 
general properties of the solutions of this equation, assuming that
the eigenvalue ${\cal E}^2$ is known.
First, if $x$ is a solution, also $1/x$ is a solution. Moreover, 
if $x$ is a complex solution, also its complex conjugate 
$\bar{x}$ is a solution. These two
properties allow us to classify the solutions of the equation into
four classes:
\begin{itemize}
\item[i)] All solutions are real, with $|x| \not=1$; they 
can be written as $x = p$, $q$, $1/q$, $1/p$, where 
$p$ and $q$ are real numbers satisfying $|p|,|q|>1$.
\item[ii)] There are four complex solutions with $|x|\not=1$. They can be 
parametrized as $x = p e^{i \phi}$,
$p e^{-i\phi}$, $e^{i\phi}/p$, $e^{-i\phi}/p$, where $p$ is a real 
positive number with $p>1$.
\item[iii)] There are two real solutions with $|x| \not=1$ 
and two complex solutions 
with $|x| = 1$. They can be parametrized as 
$x = p, 1/p, e^{i \phi}, e^{-i\phi}$, where $p$ is a real number 
with $|p|>1$.
\item[iv)] There are four complex solutions with $|x|=1$ that can be 
parametrized as $(0\le \phi_1,\phi_2 \le \pi)$ 
$x = e^{i\phi_1}, e^{i\phi_2}, e^{-i\phi_1}, e^{-i\phi_2}$. 
\end{itemize}
The solutions $x_i$ provide the two quantities $x_1$ and $x_2$, that we 
use to parametrize the eigenvectors of $\hat C$. The equations as 
well as the parametrization of the eigenvectors are 
invariant under $x_1 \to 1/x_1$, $x_2 \to 1/x_2$ and under the 
exchange of $x_1$ and $x_2$, and thus any pair of solutions that are not 
related by the inversion symmetry ($x\to 1/x$) 
can be identified with $x_1$ and $x_2$. To unambiguously 
define the two parameters, we define $x_1$ and $x_2$ such that 
$|x_1|,|x_2|\ge 1$ and $|x_1|\ge |x_2|$.  
More precisely, let us assume that the parameters $p$, $q$, $\phi$, $\phi_1$
and $\phi_2$ that we have used above to parametrize the solutions satisfy 
$|p| > |q| > 1$, $0\le \phi,\phi_1,\phi_2\le \pi$, and $\phi_1>\phi_2$. 
Then, we define:
\begin{itemize}
\item[]Solutions i): $x_1 = p$, $x_2 = q$.
\item[]Solutions ii): $x_1 = p e^{i\phi}$, $x_2 = p e^{-i\phi}$.
\item[]Solutions iii): $x_1 = p$, $x_2 = e^{i\phi}$.
\item[]Solutions iv): $x_1 = e^{i\phi_1}$, $x_2 = e^{i\phi_2}$.
\end{itemize}
In the following we will always assume that $x_1$ and $x_2$ have been determined
as discussed here.

As a first step of our analysis, we perform a numerical 
study, with the purpose of determining $x_1$ and $x_2$ 
in the infinite-chain limit, for given model parameters 
$g$, $\gamma$, and $J_0$. 
For this purpose, we determine the two lowest
eigenvalues ${\cal E}_1^2$ and ${\cal E}_2^2$ of 
the matrix $\hat{C}$ for a given size 
$L$ using a standard numerical algorithm and then we solve the equation 
$\varepsilon(x)^2 = {\cal E}_i^2$. This procedure 
allows us to compute $x_1(L)$ and $x_2(L)$ for the two levels.
We repeat the procedure for several values of $L$ (typically,
$L$ varies from 20 up to 100-300) and then
we analyze the size behavior of the solutions.
For some parameter values we observe a very fast convergence: 
Within errors, the estimates of ${\cal E}_1$ and ${\cal E}_2$ 
are approximately the same
and do not depend on $L$ for $L\gtrsim 100$.
In this case, we take the results for $L=100$ as 
the infinite-chain estimates of $x_1$ and $x_2$.
This type of behavior occurs when the parameters satisfy $|x_1| > 1$ and 
$|x_2| > 1$, i.e., when the 
solutions are of type (i) or (ii).
For some other parameter values, instead, 
${\cal E}_1$ and ${\cal E}_2$ show a significant size dependence. 
Size corrections apparently decay as an inverse power of $L$. Therefore,
if $x_i(L)$ is real, 
it is extrapolated to $a_0 + a_1/L + a_2/L^2$. If $x_i(L)$ is complex,
we separately extrapolate $|x_i(L)|$ and the corresponding phase
to $a_0 + a_1/L + a_2/L^2$. 
As expected (for $g < 1$, the ground state is degenerate in the infinite-size 
limit), the extrapolated values for the two levels are approximately
the same. This type of convergence occurs when the solutions are of type (iii) 
or (iv). 

As an example, in Fig.~\ref{x1x2-genericJ0gamma} we report 
the infinite-length estimates of
$|x_1|$ and $|x_2|$ as a function of $g$ for different values of $J_0$ and 
$\gamma$. The boundaries between the different types of 
solutions have been 
obtained using the exact results presented in the following 
sections. In the upper left panel, solutions change from type (i) to type 
(iii) as $g$ increases, while in the lower panels we go from type (iv) to type 
(iii). In the upper right panel, as $g$ increases, we observe the 
transitions (ii)$\to$ (i) $\to$ (iii).  

The analysis of the size behavior of the eigenvalues ${\cal E}_i$ suggests that 
there is a strict relation between the nature of the solutions of the 
equation $\varepsilon(x)^2 = {\cal E}_i^2$ and the phase behavior of the model. 
This is even more evident from the analysis of the gap, which shows 
that the parameter space $J_0,\gamma, g$ ($0\le g < 1$) 
can be divided into four phases. In each phase, the parameters 
$x_1$ and $x_2$ belong to one of the four classes discussed above.

If, for given Hamiltonian parameters,  $x_1$ and $x_2$ are both real 
[solutions of type (i)],
we find that the gap $\Delta = E_1 - E_0$ decreases
exponentially with the chain length. This is the behavior expected at 
a magnetic first-order transition: thus, solutions of type 
(i) characterize what we name  magnetized (M) phase. If 
the values of $x_1$ and $x_2$ for given Hamiltonian parameters are of 
type (ii), the system is also in 
a magnetized phase, since size corrections decay exponentially with $L$.
However, in this case the gap behaves as 
$\Delta = f(L) e^{-m L}$, where $f(L)$ is an oscillatory bounded function,
whose oscillations are not commensurate with the chain size. We will name 
this phase magnetized-incommensurate (MI) phase. The size behavior 
changes at phase points where  $|x_2| = 1$, i.e., 
if $x_1$ and $x_2$ are solutions of type 
(iii) and (iv). If $x_1$ is real [solutions of type (iii)], the 
gap scales as $1/L^2$: the corresponding phase will be named kink (K)
phase. When also $x_1$ is complex with $|x_1| = 1$ [solution is of type (iv)], 
there are additional 
oscillations and the gap decreases
as $\Delta = f(L)/L^2$ for large $L$, where $f(L)$ is an incommensurate 
oscillatory bounded function. This phase will be named 
kink-incommensurate (KI) phase. 

\begin{figure}[tbp]
\begin{center}
\begin{tabular}{ll}
\hspace{-1cm}
\epsfig{width=10truecm,angle=0,file=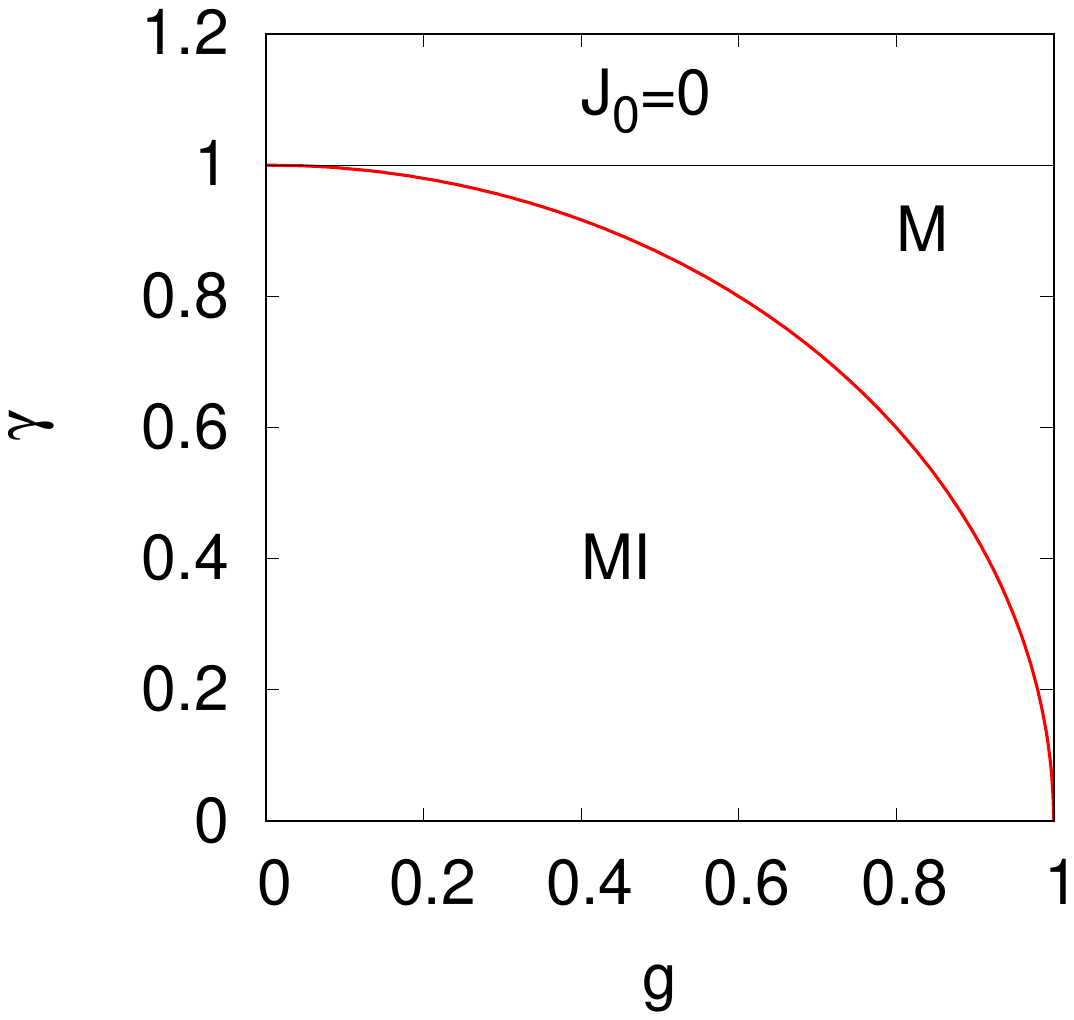} & \hspace{-3cm}
\epsfig{width=10truecm,angle=0,file=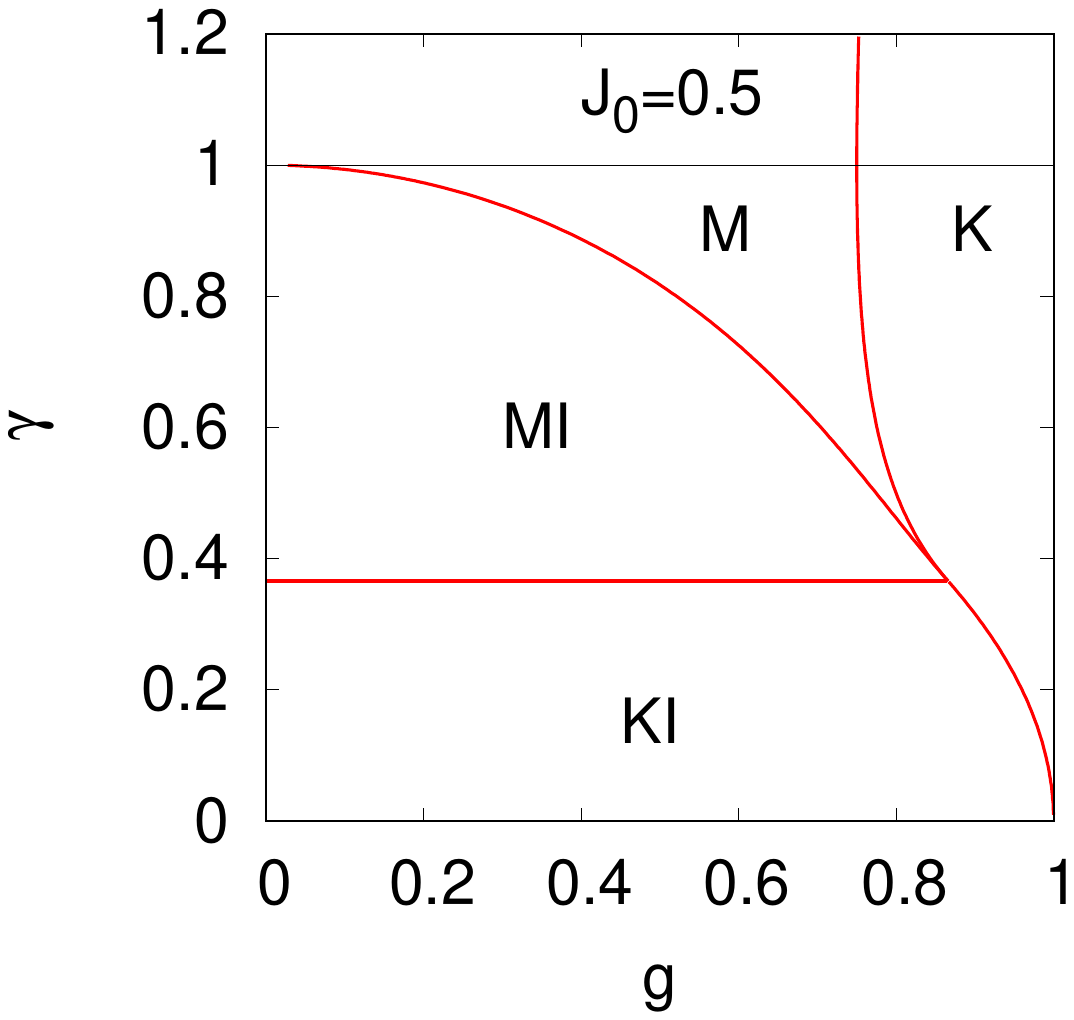} \\
\hspace{-1cm}
\epsfig{width=10truecm,angle=0,file=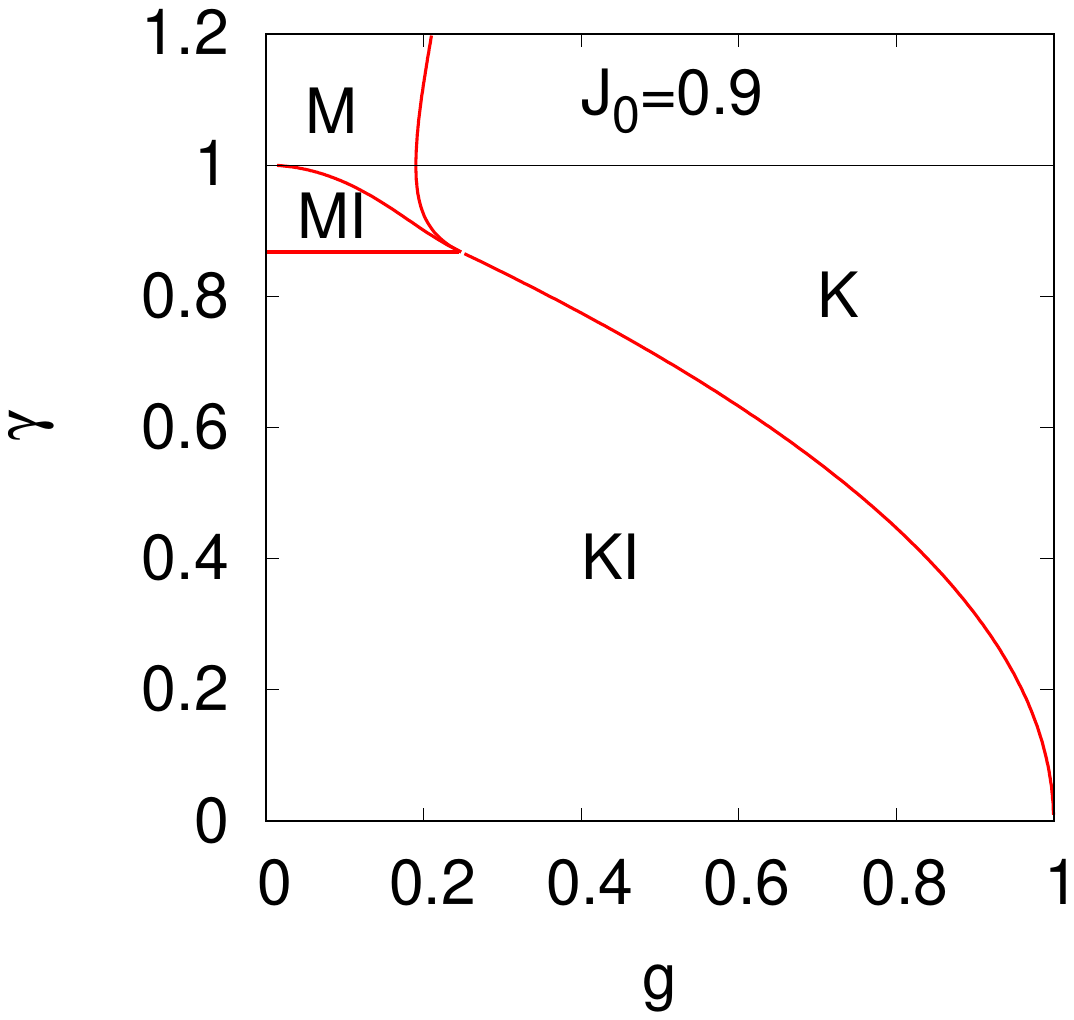} & \hspace{-3cm}
\epsfig{width=10truecm,angle=0,file=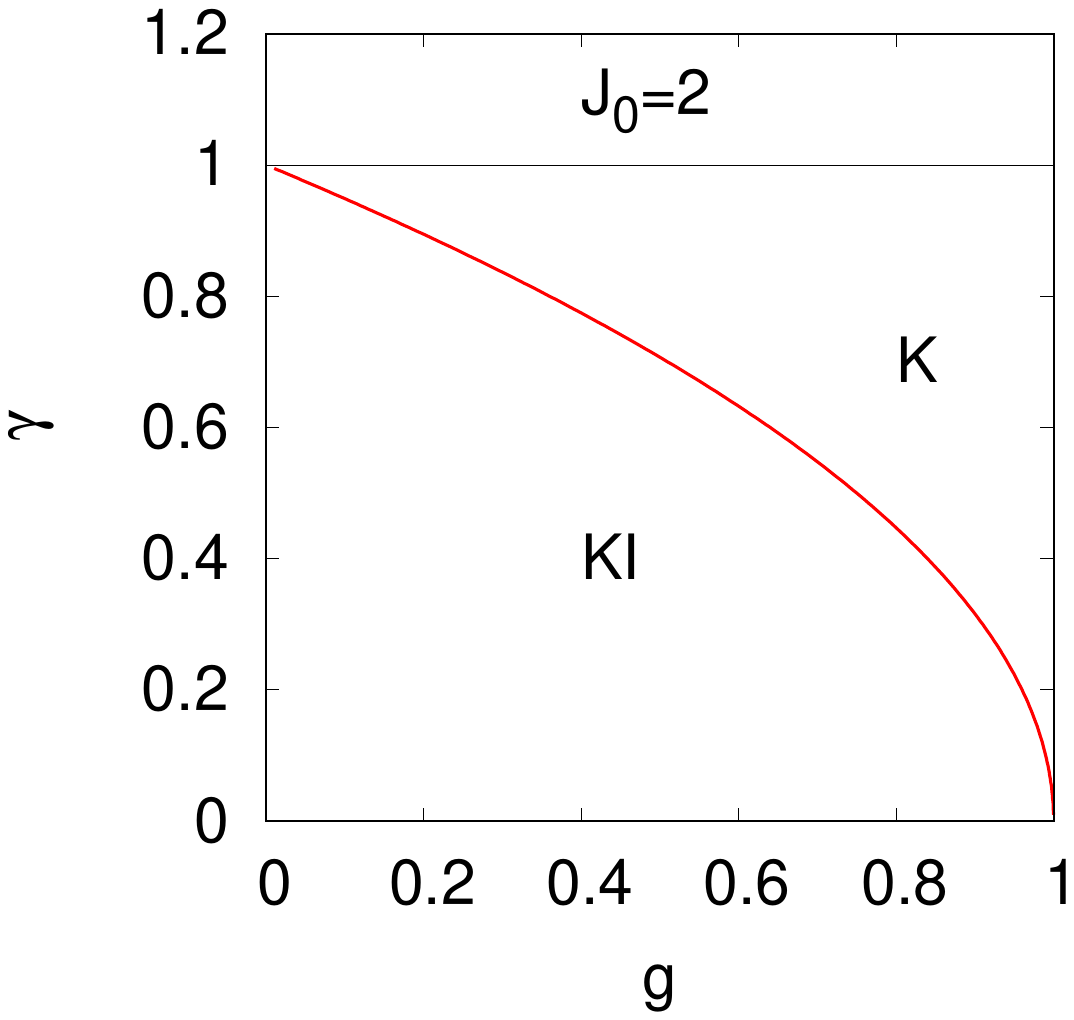} \\
\end{tabular}
\end{center}
\vskip-5mm
\caption{Phase diagram in the $g$-$\gamma$ plane for 
fixed values of $J_0$ and positive values of $\gamma$: 
behavior for $J_0=0$ (top left), 
$J_0=0.5$ (top right), $J_0=0.9$ (bottom left), 
and $J_0=2$ (bottom right). 
}
\label{Phasediagrams}
\end{figure}

\begin{figure}[tbp]
\begin{center}
\begin{tabular}{ll}
\hspace{-1cm}
\epsfig{width=10truecm,angle=0,file=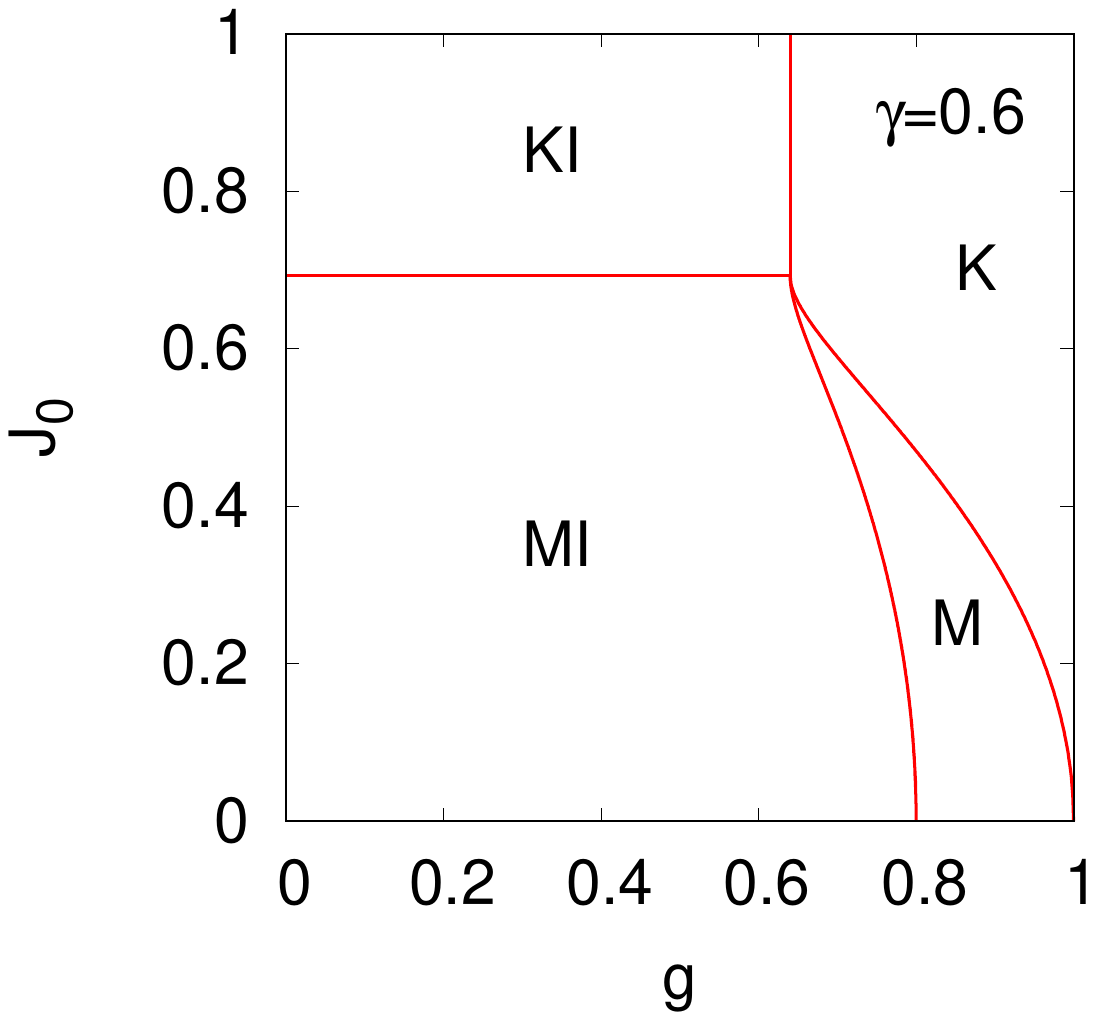} & \hspace{-3cm}
\epsfig{width=10truecm,angle=0,file=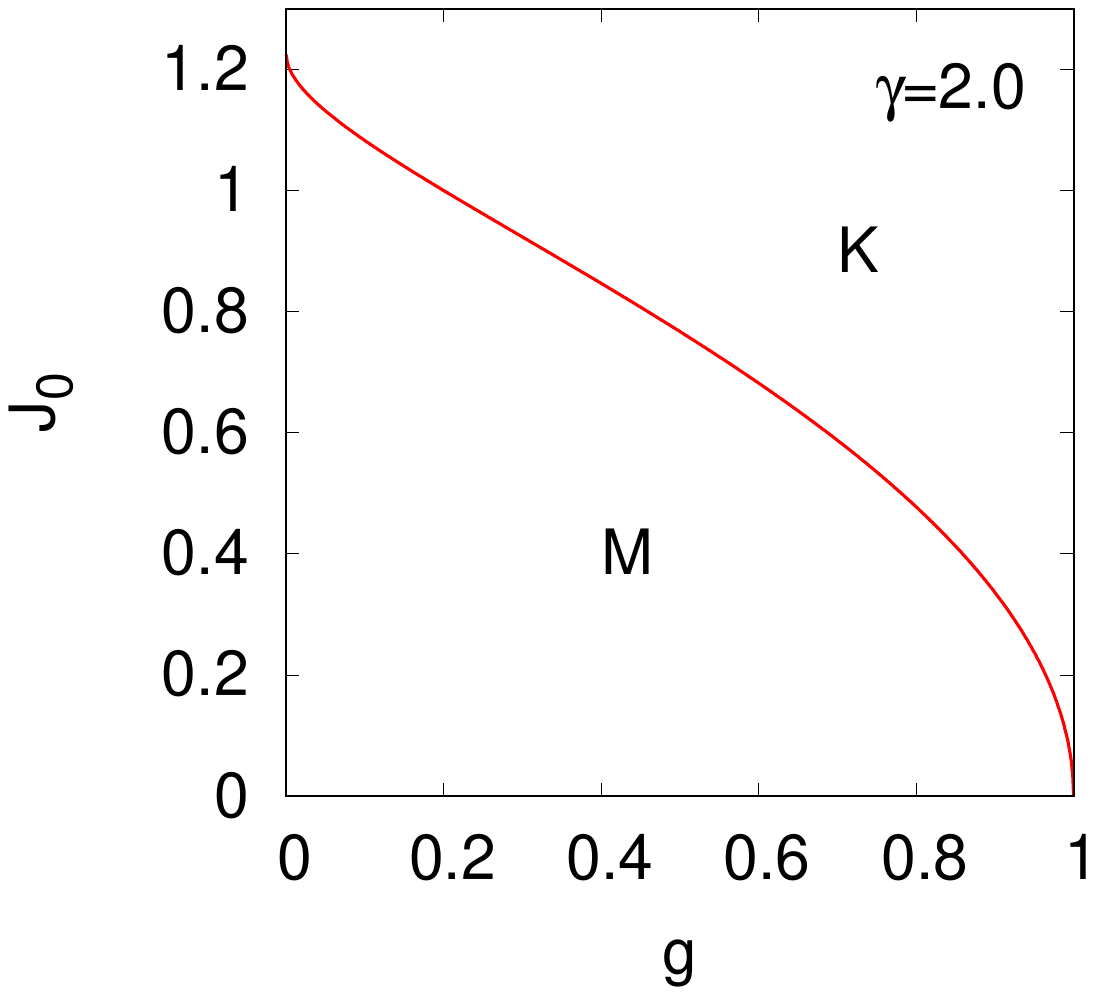} 
\end{tabular}
\end{center}
\vskip-1cm
\caption{Phase diagram in the $g$-$J_0$ plane for 
fixed values of $\gamma$:
behavior for $\gamma=0.6$ (left) and 
for $\gamma=2.0$ (right).
}
\label{Phasediagramsgamma}
\end{figure}

\begin{figure}[tbp]
\begin{center}
\begin{tabular}{c}
\epsfig{width=10truecm,angle=0,file=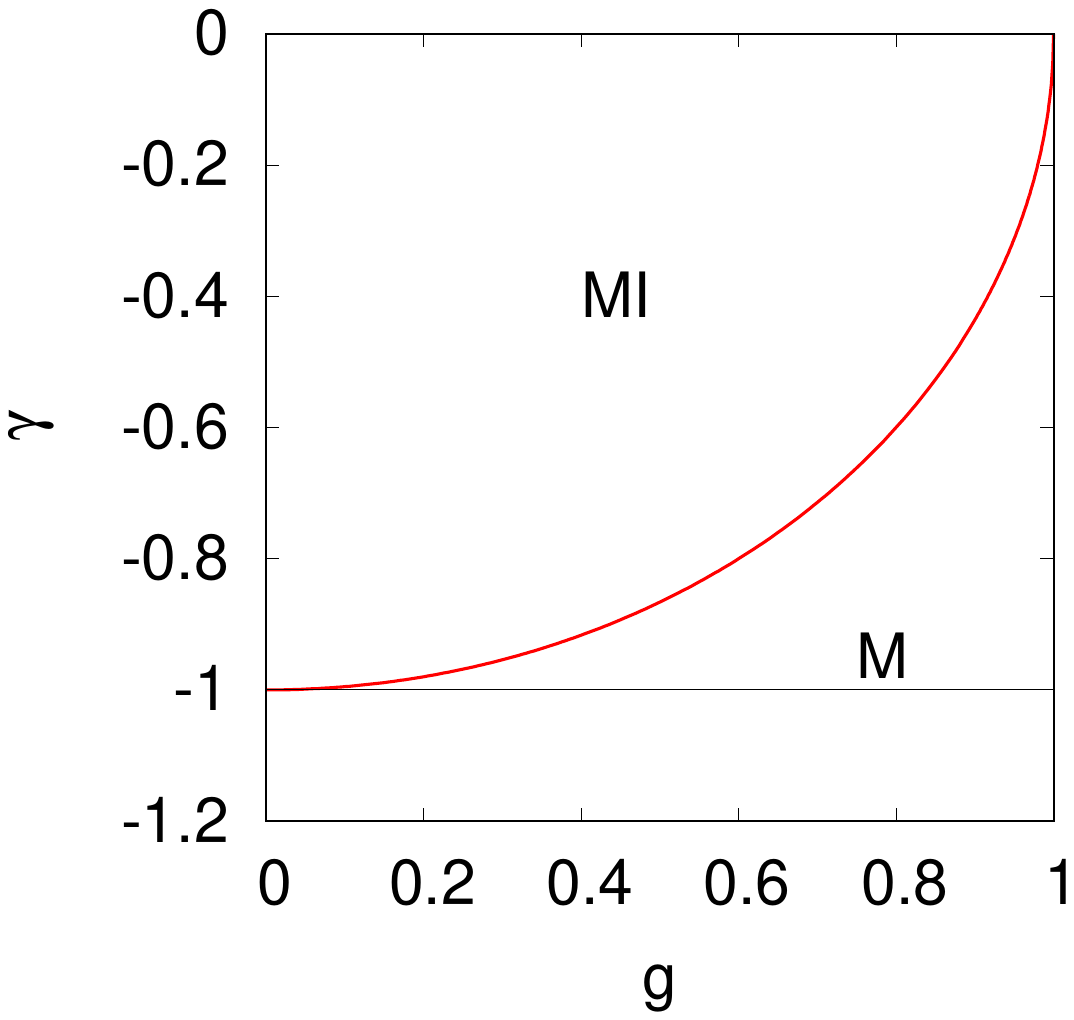} \\
\end{tabular}
\end{center}
\vskip-5mm
\caption{
Phase diagram in the $g$-$\gamma$ plane for $\gamma < 0$. 
The phase diagram does not depend on $J_0$.
}
\label{Phasediagram-gammaneg}
\end{figure}

In Figs.~\ref{Phasediagrams}, \ref{Phasediagramsgamma}, 
and \ref{Phasediagram-gammaneg} 
we anticipate the phase diagrams obtained from the analysis. 
For fixed values of $J_0$, there are three
possible different regimes that correspond to $0<J_0 < 1$ and $\gamma > 0$, 
to $J_0 \ge 1$ and $\gamma > 0$, and to $\gamma < 0$---in this last case the
value of $J_0$ is irrelevant. In the first case, we observe all four possible 
phases (see the panels for $J_0=0.5$ and $J_0 = 0.9$ in 
Fig.~\ref{Phasediagrams}). 
For small values of $\gamma$, the system is in the KI phase for $g$ small
and in the K phase for $g$ close to 1. As $\gamma$ increases the 
KI phase is replaced by the MI (for small $g$ and only up to $\gamma =1$) 
and by the M (intermediate values of 
$g$) phase.  The behavior changes as $J_0$ increases. Indeed,
the M and the MI phases shrink and 
disappear for $J_0 = 1$. For larger values of $J_0$, the phase diagram 
is independent of $J_0$, with two phases, the KI phase (in the region  
$\gamma < 1$) and the K phase: see the panel for $J_0=2$ in 
Fig.~\ref{Phasediagrams}.  In Fig.~\ref{Phasediagramsgamma} we report the 
phase diagrams for two fixed values of $\gamma$. For $\gamma = 2$, the behavior 
is analogous to that observed in the Ising chain, with a magnetized phase for 
small boundary fields and a kink phase for large values. 
For $\gamma = 0.6$, all four different phases appear.

The behavior for $\gamma < 0$ is independent of $J_0$,
see Fig.~\ref{Phasediagram-gammaneg}, 
and is the same as in the case of periodic boundary conditions: the boundary
between the MI and the M phase satisfies the equation
$g^2 + \gamma^2 = 1$. The same phase 
behavior is observed for $J_0 = 0$ (open boundary conditions). This result is quite 
obvious for the Ising chain with $\gamma = -1$. Indeed, for this value of $\gamma$ 
the system orders ferromagnetically in the $y$ direction, irrespective of the 
boundary fields that instead point in the $x$ direction.

\section{The magnetized phases}
\label{sec6}

We wish now to characterize the magnetized phases. 
In the infinite-size limit we expect two degenerate solutions
of the eigenvalue equation. No degeneracy is present for finite $L$
and therefore 
we expect two low-lying states with an exponentially small gap. 
In Sec.~\ref{sec6.1} we determine the
values of the parameters $x_1$ and $x_2$ that correspond to the
two degenerate ground-state eigenvectors 
(details are given in \ref{App.magnetized}), 
while in Sec.~\ref{sec6.2} we 
determine for which values of the parameters the system is in 
the magnetized phases (M and MI phases, as discussed in Sec.~\ref{sec5}).

\subsection{Determination of the parameters $x_1$ and $x_2$} 
\label{sec6.1}

In this section, we compute $x_1$ and $x_2$ for the ground state 
in the infinite-chain limit.
As discussed in Sec.~\ref{sec5},
in the magnetized phases $|x_1|$ and $|x_2|$ are both larger than $1$. 
Therefore, for large system sizes, the quantities
$x_1^{-L}$ and $x_2^{-L}$ are exponentially small and thus the leading 
behavior is obtained by neglecting these exponential 
terms. The equations (\ref{eq:foureqs})
are analyzed in \ref{App.magnetized}. For $\gamma > 0$, $x_1$ and $x_2$
are solutions of 
\begin{eqnarray}
  x_1 x_2 &=& {(1 + \gamma)^2 - 4 J_0^2 \over 1 - \gamma^2}  \; ,
\label{magn_eq2} \\
  x_1 + x_2 &=& {2 g [(1 + \gamma)^2 - 4 J_0^2] \over 
      (1 - \gamma^2) (1 + \gamma - 2 J_0^2)}.
\label{magn_eq3}
\end{eqnarray}
On the other hand, for $\gamma < 0$, $x_1$ and $x_2$ both satisfy 
$f_1(x_i) f_1(1/x_i) = 0$, where 
the function $f_1(x)$ is defined as
\begin{equation}
f_1(x) = (1 + \gamma) x^2 - 2 g x + (1 - \gamma) = 0.
\end{equation}
If the solutions are both real and 
larger than 1 in absolute value, the system is in 
the M phase for the given set of parameters. If they are complex conjugate 
with $|x_1| = |x_2| > 1$, the system is in the MI phase. 
    
It is important to note that, for each set of Hamiltonian parameters,
we obtain two independent eigenvectors, as expected in a magnetized phase.
In a finite volume the degeneracy is lifted. The splitting of the two
degenerate levels should be due to the terms of order $x_1^{-L}$ and $x_2^{-L}$
that we have neglected in the computation of the infinite-size behavior.
Since $|x_2| \le |x_1|$, 
the gap should scale as $|x_2|^{-L}$, with incommensurate 
oscillations if $x_2$ is complex. 

\begin{figure}[tbp]
\begin{center}
\begin{tabular}{cc}
\epsfig{width=8truecm,angle=0,file=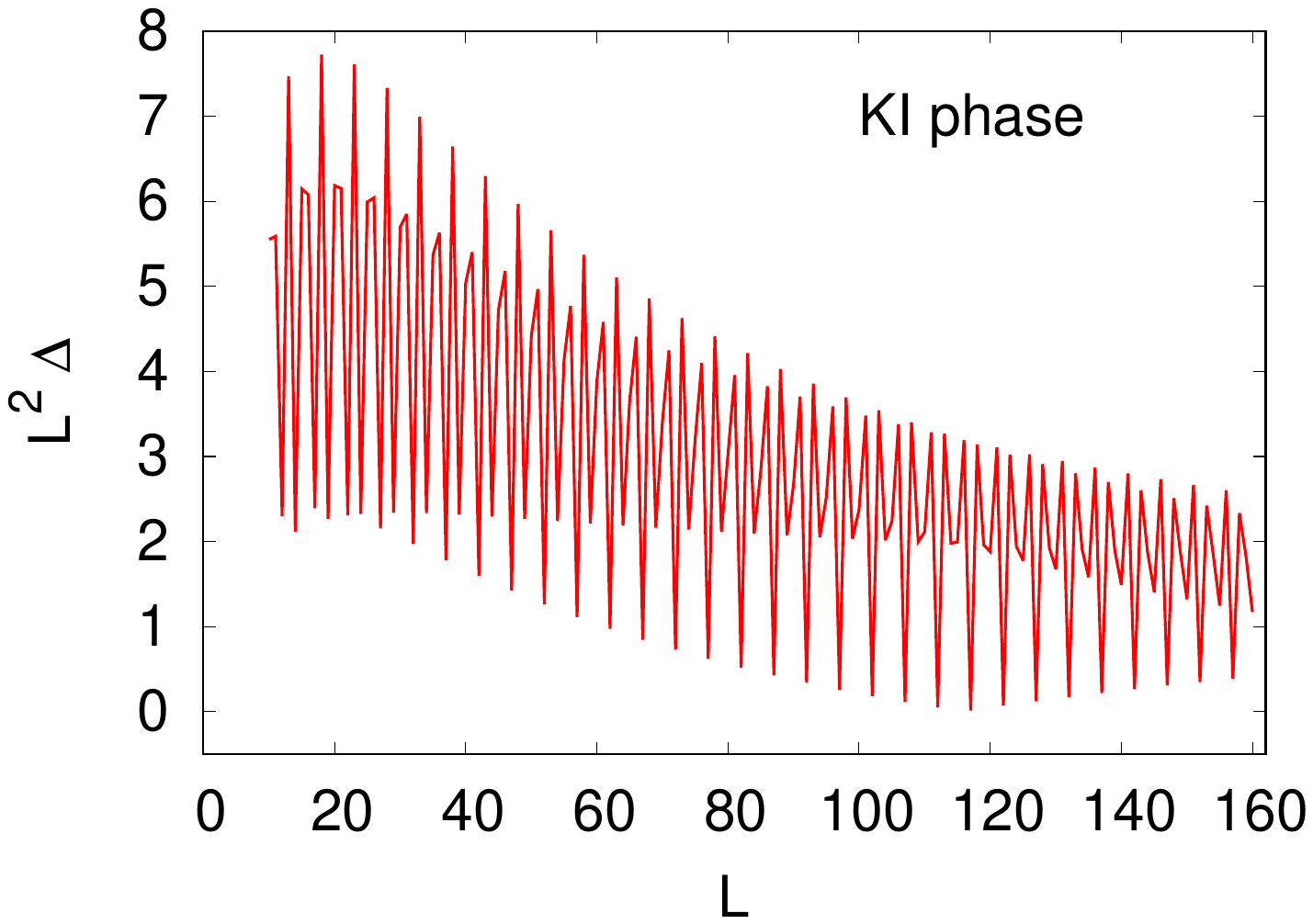} & \hspace{-1cm}
\epsfig{width=8truecm,angle=0,file=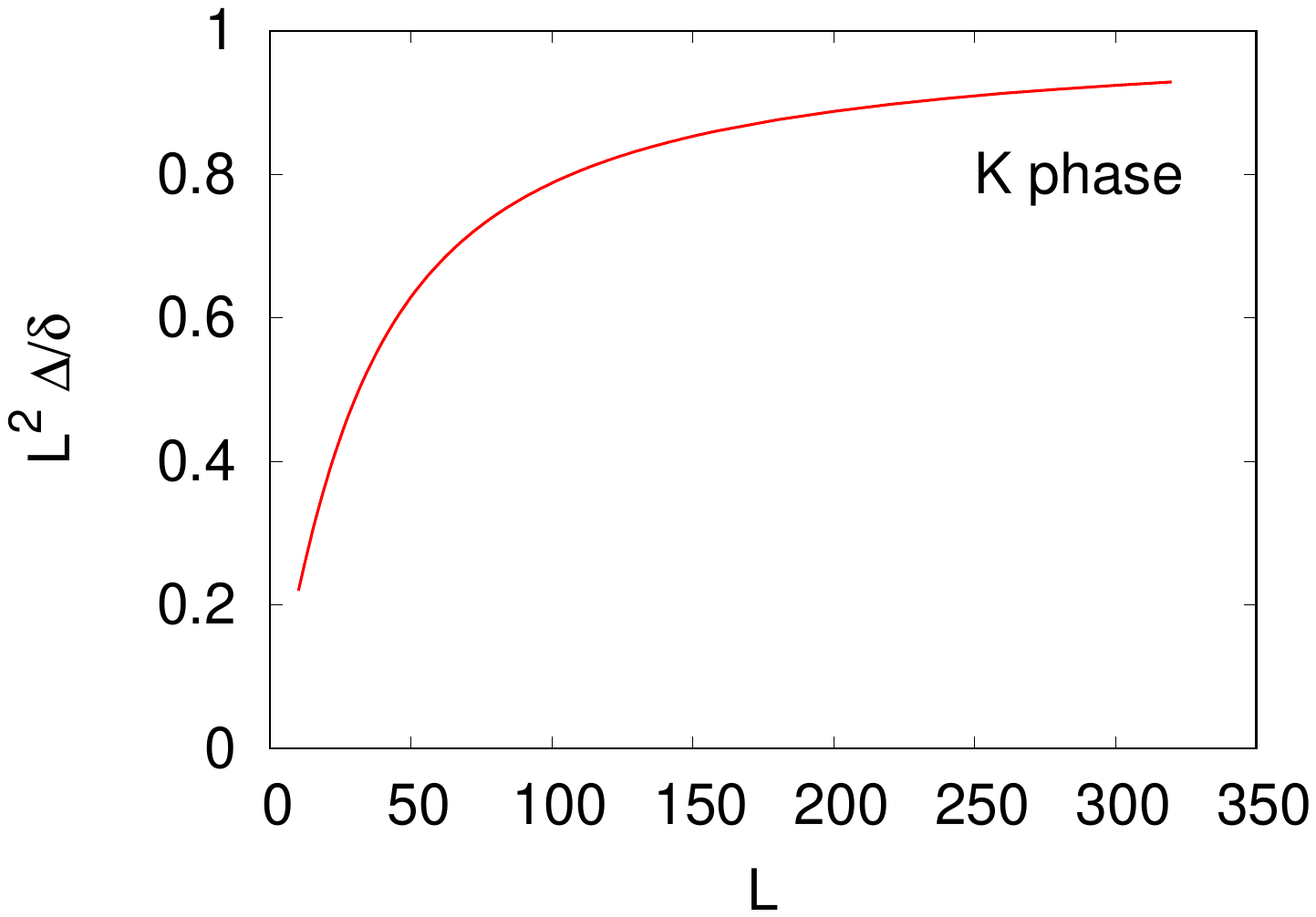} \\
\epsfig{width=8truecm,angle=0,file=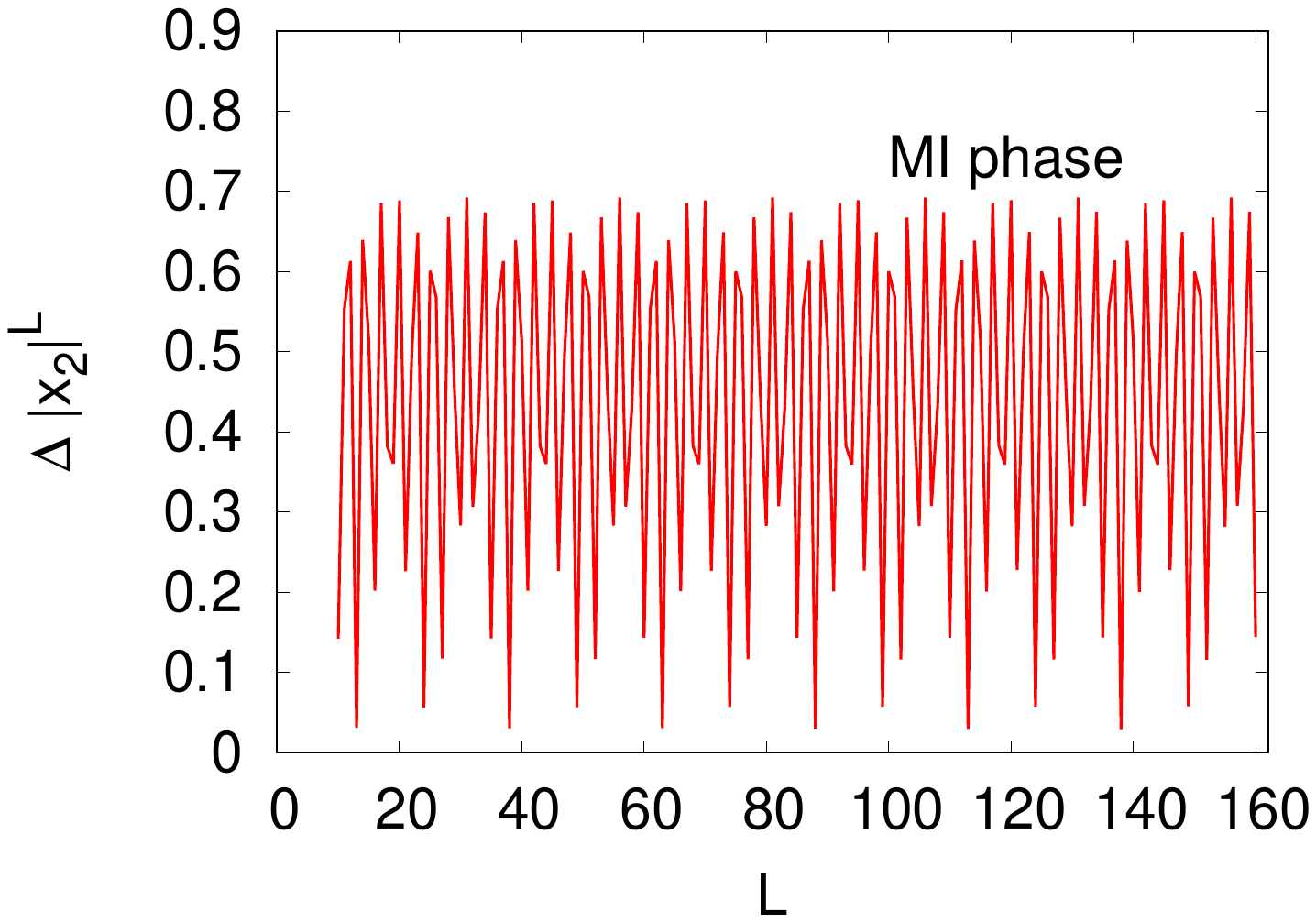} & \hspace{-1cm}
\epsfig{width=8truecm,angle=0,file=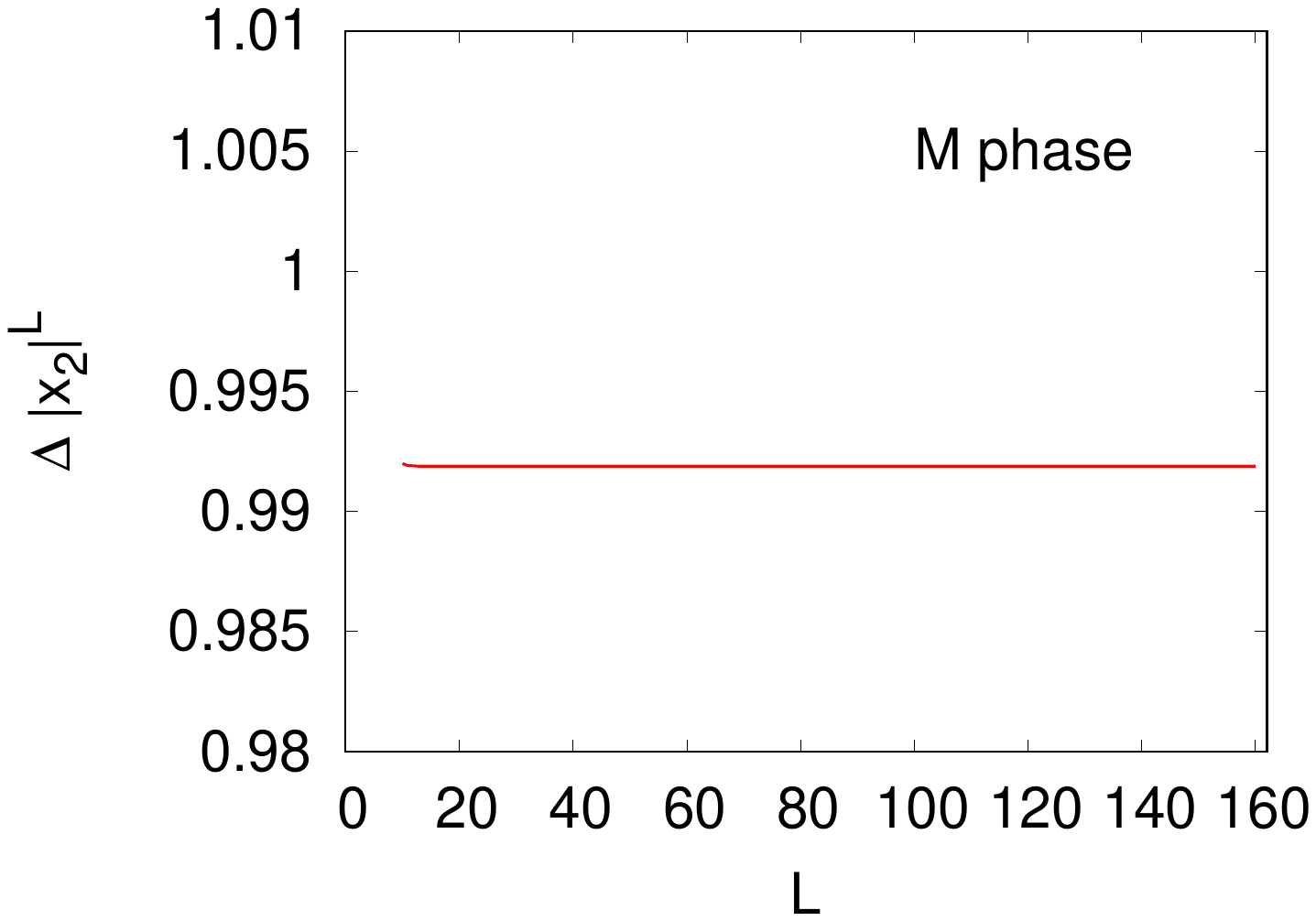} \\
\end{tabular}
\end{center}
\vskip-5mm
\caption{Rescaled gap for different sizes in the four phases.
Top left: $g=0.3$, $\gamma = 0.2$ (KI phase); 
Top right: $g=0.9$, $\gamma = 0.6$ (K phase); 
Bottom left: $g=0.3$, $\gamma = 0.6$ (MI phase); 
Bottom right: $g=0.3$, $\gamma = 1.2$ (M phase). In all cases 
$J_0 = 0.5$. The gap $\Delta$ has been determined numerically for several 
values of $L$, diagonalizing the matrix $\hat{C}$. 
We use the exact results for $x_2$ and $\delta$. We have
[see Eqs.~(\ref{magn_eq2}) and (\ref{magn_eq3})]
$|x_2| = 1.5612495$ for $g=0.3$, $\gamma = 0.6$ (MI phase),
$x_2 =   1.7914396$ for $g=0.3$, $\gamma = 1.2$ (M phase),
while, for $g=0.9$ and $\gamma = 0.6$ (K phase), the gap behaves as 
$\delta/L^2$ with $\delta = 76.9829$ [see Eq.~(\ref{DeltaK})].}
\label{RescaledGap}
\end{figure}

In Fig.~\ref{RescaledGap} we show the rescaled gap for some points that
correspond to the four different phases. The results for the MI and M phases
(see the lower panels)
are rescaled by $|x_2|^L$ and $x_2^L$, respectively, where $x_2$ 
has been determined by using Eqs.~(\ref{magn_eq2}) and (\ref{magn_eq3}). 
In the M case, $x_2^L \Delta(L)$ 
is apparently independent of $L$ for $L\ge 20$, 
indicating that $\Delta(L) \approx a x_2^{-L}$ with negligible 
size corrections (they are expected to be of order $x_2^{2 L}$ and $x_1^L$).
In the MI case,
the rescaled data are constant on average, but show an oscillatory behavior:
apparently we have $|x_2|^L \Delta(L) \approx f(L)$ where $f(L)$ is a bounded
oscillating function, so that $\Delta(L) \approx f(L) |x_2|^{-L}$.

Let us note that, in the M case, $x_1$ and $x_2$ should be 
both real and their absolute value should be larger than 1. This condition
does not exclude that one or both of them is negative, smaller than $-1$. 
Numerically, we have found that $x_2$ is always positive, excluding the
possibility of oscillations of the gap with the parity of 
the size: indeed, if $x_2 < -1$ we would have 
$\Delta(L) \approx a (-1)^L |x_2|^{-L}$.  
The absence of even-odd oscillations is expected as the model is ferromagnetic. 
Instead, $x_1$ is positive or negative, depending on $\gamma$:
  $x_1$ is positive for $|\gamma| < 1$ and negative in the 
opposite case. This last result can be explained by noting that,
for  $|\gamma| > 1$, 
the Hamiltonian is the sum of a ferromagnetic dominant hopping term 
and of an antiferromagnetic subdominant hopping term. Apparently, the latter
term induces subdominant even-odd oscillations that decrease as 
$(-1)^L |x_1|^L$.

Finally, let us discuss the limits $\gamma \to \pm 1$, in which the 
XY chain becomes the simpler Ising chain. For $\gamma \to 1$, we have 
\begin{equation}
 x_1 \approx  {2g\over 1- \gamma} \qquad x_2 = {1 - J_0^2\over g},
\end{equation} 
As expected, $x_1$ and $x_2$ are real for $\gamma$ close to 1, in 
agreement with the Ising results: no MI phase is observed in the Ising chain. 
Moreover, $x_1$ diverges as $\gamma\to 1$, in agreement with the 
fact that the Ising solution can be parametrized in terms of a 
single variable $x$, see Ref.~\cite{Hu2021}. 
The result for $x_2$ is in agreement with 
the result for the gap obtained in Ref.~\cite{CPV-15JSTAT}.

For $\gamma \to -1$ we obtain instead
\begin{equation}
 x_1 \approx  {2g\over 1+ \gamma} \qquad x_2 = {1\over g},
\end{equation} 
As before, $x_1$ diverges on the Ising line. Here $x_2$ is independent of
$J_0$, and therefore the gap scales as for open boundary conditions 
($J_0=0$).

\subsection{Boundaries of the magnetized phases} 
\label{sec6.2}

We are now in the position to discuss the boundaries of the magnetized
phases. We first discuss the behavior for $\gamma \ge 0$. In this case,
the values of $x_1$ and $x_2$ are determined by using Eqs.~(\ref{magn_eq2})
and (\ref{magn_eq3}). Let us first determine the boundary 
between the M and the MI phase, Since, in the MI phase $x_1$ and $x_2$ 
are complex numbers satisfying $|x_1| = |x_2|$, while in the M phase they 
are real, on the boundary $x_1$ and $x_2$ should be real satisfying 
$x_1 = \pm x_2$. As we have discussed above, the MI phase 
lies in the region $J_0 < 1$ and $\gamma < 1$. In this range of parameters
both $x_1$ and $x_2$ are positive and therefore the boundary is 
defined by $x_1 = x_2$.
If we set $x_1 = x_2$ in Eqs.~(\ref{magn_eq2})
and (\ref{magn_eq3}) and solve for $g$  (assuming $J_0 < 1$ and $\gamma < 1$)
we obtain
\begin{equation}
g_{M/MI} = (1 + \gamma - 2 J_0^2) 
    \sqrt{1-\gamma^2 \over (1 + \gamma)^2 - 4 J_0^2}  \; .
\label{gMMI}
\end{equation}
At the boundary we have 
\begin{equation}
x_{1,M/MI} = x_{2,M/MI} = 
     \sqrt{(1 + \gamma)^2 - 4 J_0^2 \over 1 - \gamma^2} . 
\end{equation}
Since $J_0 < 1$, $g_{M/MI}$ decreases
as $\gamma$ increases. Moreover, we have $g_{M/MI} = 0$ for $\gamma = 1$. 
Note that we have $g_{M/MI} = \sqrt{1 - \gamma^2}$ for $J_0 = 0$, which is 
the result that also holds for periodic boundary conditions.

To derive the boundary between the KI phase and the MI phase, let us note 
that $x_1 = p e^{i\phi}$ and $x_2 = p e^{-i\phi}$ 
in the MI phase,
and $x_1 = e^{i\phi_1}$ and $x_2 = e^{i\phi_2}$ in the KI phase.
Thus, as we move in the MI phase towards the boundary, the parameter
$p$ converges to $p = 1$. Therefore, the boundary is characterized
by the condition $x_1 x_2 = 1$.
Eq.~(\ref{magn_eq2}) then gives
\begin{equation}
\gamma_{KI/MI} = {1\over 2} \left(
    \sqrt{1 + 8 J_0^2} - 1\right),
\end{equation}
which is independent of $g$.  Note that $\gamma_{KI/MI}$ increases with 
increasing $J_0$ and that $\gamma_{KI/MI} = 0,1$ for 
$J_0 = 0,1$, respectively. Moreover,
$x_{1,M/MI}=x_{2,M/MI}$ 
is larger than 1 only for $\gamma > \gamma_{KI/MI}$, so that 
the MI phase lies in the region $\gamma > \gamma_{KI/MI}$
(see the phase diagrams for $J_0=0.5$ and 0.9 in Fig.~\ref{Phasediagrams}). 

Let us finally determine the boundary between the
K phase ($x_1 > 1$ and $|x_2|=1$) and the 
M phase ($|x_1|,|x_2| > 1$), which only exists for 
$J_0 < 1$, see the plots shown in Fig.~\ref{Phasediagrams}.
As we have already discussed, in the M phase $x_2$ is always positive and 
therefore, the boundary is defined by the condition
$x_2 = 1$. Substituting $x_2 = 1$
in Eqs.~(\ref{magn_eq2}) and (\ref{magn_eq3}) 
we obtain 
\begin{equation}
g_{K/M} = {(1 + \gamma - 2 J_0^2)^2 \over 
          (1 + \gamma)^2 - 4 J_0^2} , 
\label{gKM}
\end{equation}
with 
\begin{equation}
x_{1,K/M} = {(1 + \gamma)^2 - 4 J_0^2 \over 1 - \gamma^2}.
\label{xKM}
\end{equation}
For $\gamma = 1$, we obtain $g_{K/M} = 1 - J_0^2$, in agreement with the 
results of Ref.~\cite{CPV-15JSTAT}. 
For $J_0 < 1$, the two conditions, 
(i) $g_{K/M}$ is real and lies in $[0,1]$, and (ii) $x_{1,K/M} > 1$, 
are satisfied only for $\gamma > \gamma_{KI/MI}$. The boundary 
therefore lies in the region $\gamma > \gamma_{KI/MI}$. 
Finally, let us note that,
for fixed $J_0 < 1$, $g_{K/M}$ decreases
with increasing $\gamma$,
reaches its minimum at $\gamma = 1$, and then increases, converging 
to 1 for $\gamma \to \infty$ (see the phase diagrams for 
$J_0 = 0.5$ and $J_0 = 0.9$ in Fig.~\ref{Phasediagrams}). 
Correspondingly, $x_{1,K/M}$ is an increasing positive function of $\gamma$ for 
$\gamma < 1$, while it is a decreasing negative function of $\gamma$ for 
$\gamma > 1$.

The three boundaries meet along a multicritical (MC) line
characterized by the condition $x_1 = x_2 = 1$.  We obtain 
\begin{equation}
\gamma_{MC} = \gamma_{KI/MI} \qquad 
g_{MC} = {1\over2} \left(\sqrt{1 + 8 J_0^2} + 1 - 4 J_0^2\right) = 
1 - \gamma_{MC}^2.
\label{magn_MCpoint}
\end{equation}
The multicritical line only exists for $J_0 < 1$---for $J_0 > 1$ we 
indeed obtain $g_{MC} < 0$---as the KI and MI phases.
Note that $g_{MC}$ is a 
decreasing function of $J_0$, satisfying $g_{MC} = 1,0$ for $J_0 =0$ and 1, 

The behavior for $\gamma < 0$ is much simpler. In this case, $x_1$ and $x_2$ 
are solutions of $f_1(x_1) = f_1(x_2) = 0$. They  satisfy 
$|x_1| > 1$ and $|x_2| > 1$, and therefore 
there are only the 
M and the MI phases. For $\gamma \le -1$, all points belong to 
the M phase, while, for $-1<\gamma < 0$, systems with $g < \hat{g}_{M/MI}$
are in the MI phase and systems with $g > \hat{g}_{M/MI}$ are in the M 
phase. The boundary is specified by
\begin{equation}
   \hat{g}_{M/MI} = \sqrt{1 - \gamma^2} \qquad 
   x_{1,M/MI} = x_{2,M/MI} = {g\over 1 + \gamma},
\end{equation}
The corresponding phase 
diagram is reported in Fig.~\ref{Phasediagram-gammaneg}.

It is interesting to compare our results with the expressions 
reported in Ref.~\cite{Hu2023}. They used an approximate Ansatz 
to solve the eigenvalue equations, which is correct, in the infinite-chain
limit, only in the magnetized phases.  This approximation allowed 
them to correctly identify the boundary between the M phase and the MI
phase, Eq.~(\ref{gMMI}), as well as the position of the 
multicritical line, Eq.~(\ref{magn_MCpoint}).

\section{The kink phase} \label{sec7}

The kink phase has been extensively discussed in Ref.~\cite{CPV-15JSTAT}
for the Ising case $\gamma = 1$.
For a finite chain of length $L$ the
energies ${\cal E}_n$ of the lowest levels 
(the ``kink" states) are proportional to $n^2/L^2$ and 
thus in the infinite-length limit  there is an infinite number of 
degenerate states. As we shall see, 
the same result holds for the XY model.

In the kink phase the relevant solutions are those of type (iii), 
i.e., $x_1$ is real and larger than 1 in absolute value, and $x_2$ is 
a complex number that satisfies $|x_2|=1$ and that 
can therefore be written as $x_2 = e^{i\phi}$. 
A detailed analysis of the equations in Eq.~(\ref{eq:foureqs}) is 
reported in \ref{App.kink}. It turns out that, if we only consider the 
low-lying energy states, we have for large $L$
\begin{equation}
   x_1 = x_{10} + O(L^{-2}) \qquad 
   \phi = {\theta_1\over L} + O(L^{-2}).
\label{x1x2-K}
\end{equation}
where 
\begin{equation}
   x_1 = x_{10} \equiv {1\over 1 - \gamma^2} 
   [2 \sqrt{g (\gamma^2 + g - 1}) + \gamma^2 + 2 g - 1].
\label{kink-eq1}
\end{equation}
The phase $\theta_1$ depends on the level. For the $n$-th level of the 
matrix $\hat{C}$ we have 
$\theta_{1,n} = n \pi$. Thus, the low-energy 
eigenvectors of $\hat{C}$ are superpositions of a ``localized" 
excitation (the contribution depending on $x_1$) and of a delocalized
excitation of momentum $q = \phi = n \pi/L$, as for $\gamma = 1$.

Since the lowest lying state 
corresponds to $x_{2,1} = e^{i \pi/L}$, 
while the first excited state corresponds to 
$x_{2,2} = e^{ 2 i \pi/L}$ 
(with corrections of order $1/L^2$), the gap in the K phase is 
\begin{equation}
\Delta = \varepsilon(x_{2,2}) - \varepsilon(x_{2,1})
 = {3 \pi^2 (\gamma^2 + g - 1) \over 1 -g} {1\over L^2} + O(L^{-3}).
\label{DeltaK}
\end{equation}
If we set $\gamma = 1$ we reobtain the Ising-chain result of 
Ref.~\cite{CPV-15JSTAT}. 
A numerical check of the validity of Eq.~(\ref{DeltaK}) is provided 
in Fig.~\ref{RescaledGap}. For $g=0.9$, $\gamma = 0.6$, and $J_0=0.5$ 
we determine $\Delta(L)$ by diagonalizing numerically the 
matrix $\hat{C}$. Then, we report $\Delta(L) L^2/\delta$, 
where $\delta/L^2$ is the expected leading behavior computed using 
Eq.~(\ref{DeltaK}). The ratio converges to 1 as $L$ increases, 
confirming the correctness of Eq.~(\ref{DeltaK}).

In the K phase the parameter $x_1$ should be real, which is only true for 
$g > 1 - \gamma^2$. Since $x_{10} = 1$ for $g = 1 - \gamma^2$, this 
condition characterizes the boundary between the K phase 
and the KI phase, i.e., the K-KI boundary is given by
\begin{equation}
  g_{K/KI} = 1 - \gamma^2.
\label{gKKI}
\end{equation}
This surface lies in the region $\gamma < 1$ and also ends at the 
multicritical line (\ref{magn_MCpoint}), see, for example, the 
phase diagrams for $J_0 = 0.5$ and 0.9 in Fig.~\ref{Phasediagrams}. 

The K-KI boundary can also be determined by analyzing the minima of the 
dispersion relation (\ref{dispersion}), which should apply both in the K and 
the KI phase as the low-energy behavior is associated with propagating 
excitations
of {\em real} momentum $q$. In the K phase, $g > g_{K/KI}$, 
$\varepsilon(e^{iq})$ has a minimum for $ q = 0$, in agreement with 
the idea that the low-energy excitations are kink states of momentum
$q \sim 1/L$. On the other hand, in the KI phase the low-energy behavior is 
associated with propagating excitations of momenta $q=\pm q^*$ with 
$\cos q^* = g/(1 - \gamma^2)$. 
Refs.~\cite{CBFG-22,SCCFSZ-23} 
used these arguments to discuss the phase 
behavior of an antiferromagnetic XY model with 
frustrated boundary conditions. They found two different phases for 
$g > g_{K/KI}$ and $g < g_{K/KI}$, that we can identify with the 
K and KI phases 
that are present in the ferromagnetic case with OBF.

\section{The boundary between the K and the M phases}
\label{sec8}

\subsection{Behavior along the boundary}

Let us now determine the gap on the 
boundary between the M phase and the K phase. 
Eqs.~(\ref{x1x2-K}) and (\ref{kink-eq1}) hold also on the boundary, 
with $x_{10} = x_{1,K/M}$. Moreover, also in this case the 
phases $\theta_1$ are integer multiples of $\pi$, i.e., 
$\theta_1 = k \pi$.
However, while inside the K phase only positive values of $k$ are allowed,
on the boundary the value of $x_2$ corresponding to the 
state with energy ${\cal E}_1$ (the lowest eigenvalue) is
$x_2 = 1$ with exponential corrections, which 
implies $\theta_1 = 0$. Therefore, on the boundary we have 
$\theta_{1,n} = (n-1)\pi$ for the $n$-th level.
This different behavior for points
inside the K phase and on the M-K boundary is not surprising, as 
the same occurs in the Ising chain (for this model an analytic 
proof is given in Ref.~\cite{CPV-15JSTAT}).
Correspondingly, we obtain for the gap
$\Delta = \varepsilon(x_{2,1}) - \varepsilon(x_{2,0})$:
\begin{equation}
\Delta = {\pi^2 (\gamma^2 + g - 1) \over 1 -g} {1\over L^2} + O(L^{-3}),
\end{equation}
in full agreement with the Ising chain result \cite{CPV-15JSTAT}.

\subsection{Crossover behavior across the M-K boundary}

Refs.~\cite{CPV-15,CPV-15JSTAT} showed the presence of a universal 
crossover behavior across the boundary between the M and the K phase. 
We will now show that the same behavior occurs for generic values of
$\gamma$. We work with $g$ and $\gamma$ fixed as in 
Ref.~\cite{CPV-15JSTAT} and vary $J_0$ close to the boundary point 
\begin{equation}
J_{0c} = \sqrt{{1\over 2} (\gamma - g + 1 - s)} \qquad 
s = \sqrt{g (\gamma^2 + g - 1)},
\end{equation}
which is obtained by inverting Eq.~(\ref{gMMI}).

We parametrize the crossover in terms of the scaling variable 
\cite{CPV-15,CPV-15JSTAT}
\begin{equation}
 \zeta_s = R L (J_0 - J_{0c}) ,
\label{defzetas}
\end{equation}
where $R$ is a nonuniversal parameter that will be determined below. 
In terms of this variable, in the large-$L$ limit, we 
expect the scaling behavior
\begin{equation}
   { \Delta(J_0)\over \Delta(J_{0c}) } = f_\Delta(\zeta_s).
\label{Deltascaling-MK}
\end{equation}
The constant $R$ can be fixed so that 
$f_\Delta(\zeta_s)$ is universal, i.e., it is the same on the whole M-K 
boundary. In particular, we choose $R$ to obtain the 
same scaling curves determined in Ref.~\cite{CPV-15JSTAT}.
The details of the calculation are reported in \ref{App.MK}. We find 
\begin{equation}
   R = {4 g J_{0c} \over (g + s)^2} \qquad 
        s = \sqrt{g (\gamma^2 + g - 1)}.
\label{Rconst}
\end{equation}
We can compare this expression with the Ising result obtained in
Ref.~\cite{CPV-15JSTAT}. 
For $\gamma = 1$, the boundary occurs at $J_{0c} = \sqrt{1-g}$. 
Correspondingly we find $R = J_{0c}/g$, in agreement
with Ref.~\cite{CPV-15JSTAT}.

\begin{figure}[tbp]
\begin{center}
\begin{tabular}{c}
\epsfig{width=12truecm,angle=0,file=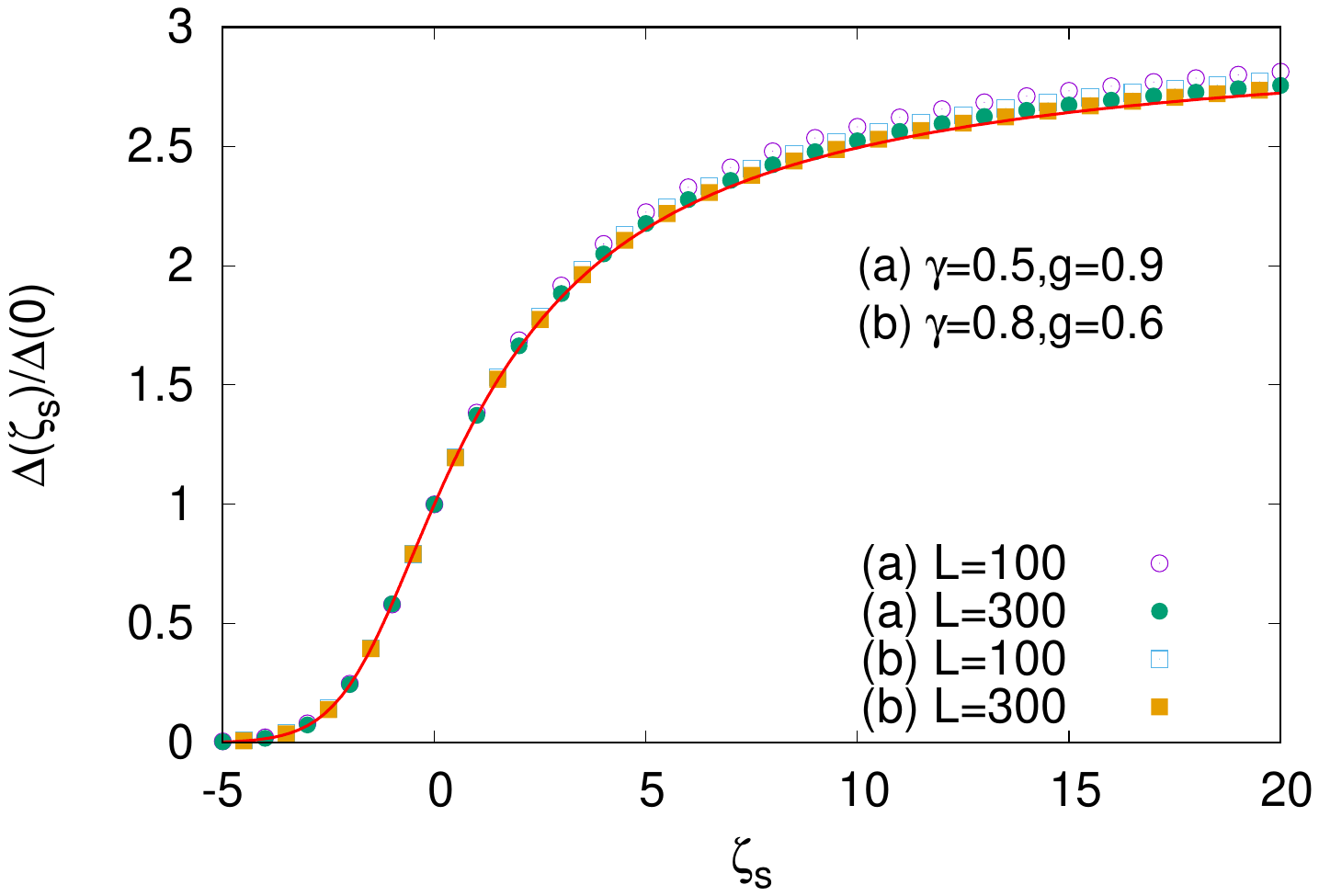}\\
\end{tabular}
\end{center}
\vskip-1cm
\caption{Ratio $\Delta(J_0)/\Delta(J_{0c})$ as a function
of $\zeta_s=RL(J_0-J_{0c})$ 
for two distinct points belonging to the M-K boundary. We used: 
(a) $J_{0c} = 0.341010$, $R = 0.764235$ for $\gamma = 0.5$ and $g=0.9$.
(b) $J_{0c} = 0.640518$, $R = 1.60235$ for $\gamma = 0.8$ and $g=0.6$.
The gap $\Delta(J_0)$ has 
been obtained by a numerical diagonalization of the matrix 
$\hat{C}$ for two different values of $L$.
The continuous line is the universal scaling curve $f_\Delta(\zeta_s)$ 
computed in Ref.~\cite{CPV-15JSTAT}.}
\label{MK-Crossover}
\end{figure}

To verify the previous calculations, 
in Fig.~\ref{MK-Crossover} we show the ratio $\Delta(J_0)/\Delta(J_{0c})$
for two different values of $L$ and two different points belonging to the 
M-K boundary (the gap is obtained by a direct diagonalization of the matrix 
$\hat{C}$). Using the value of $R$ reported in 
Eq.~(\ref{Rconst}), 
as $L$ increases the data approach the universal curve derived 
in Ref.~\cite{CPV-15JSTAT}, confirming the universality of the 
scaling behavior.

\section{The boundary between the K and the KI phases}
\label{sec9}

\subsection{Behavior along the boundary}

We now discuss the behavior along the K-KI boundary, which is 
defined by the relation 
$g = 1 - \gamma^2$, see Eq.~(\ref{gKKI}), with $\gamma > 0$. 
In the infinite-chain limit we have $x_1 = |x_2| = 1$
for the ground state. To proceed further, we 
need to determine the finite-size corrections to the previous relation.
We have first performed a finite-size numerical analysis, 
finding that $x_1$ is real and larger than
1 for finite values of $L$, while $x_2$ is a phase that converges to 1
as $L$ increases. Thus, for large values of $L$ we can write 
\begin{equation}
x_1 = 1 + {x_{11}\over L} + O(L^{-2}) \qquad
x_2 = 1 + {i \theta_{1}\over L} + O(L^{-2}),
\label{expx1x2KKI}
\end{equation}
with $x_{11}$ and $\theta_1$ strictly positive.
Inserting these expansions in the equation 
$\varepsilon(x_1)^2 = \varepsilon(x_2)^2$, we find $x_{11} = \theta_1$. 
To compute $\theta_1$, we proceed as before, 
keeping both $x_1^{-L}$ and $x_2^{-L}$,
as both quantities have a finite limit as $L\to\infty$. We consider the 
equations (\ref{eq:foureqs}), setting $c_2 = 1$. We eliminate 
$c_1$, $d_1$, and $d_2$, obtaining a single equation for $\theta_1$, 
which is then expanded in powers of $1/L$. The leading term
vanishes. Requiring the  next-to-leading term to vanish, we obtain the relation
\begin{equation}
 \cos\theta_1 \cosh\theta_1 = 1.
\label{eq-cos-cosh}
\end{equation}
This relation, which is the same for all points that belong to the 
K-KI boundary, allows us to compute $\theta_1$. There is a trivial 
solution $\theta_1 = 0$ and 
an infinite number of positive solutions that 
we indicate with $\bar{\theta}_{1,k}$, $k \ge 1$. 
The smallest positive solution $\bar{\theta}_{1,1}$ is 
\begin{equation}
\bar{\theta}_{1,1} = 4.73004\ldots
\label{theta11KKI}
\end{equation}
Solutions $\bar{\theta}_{1,n} > \bar{\theta}_{1,1}$, $n=2,3\ldots$, 
can be accurately determined using 
\begin{equation}
\bar{\theta}_{1,n} = k_n + {(-1)^n\over \cosh k_n} \qquad 
k_n = {\pi\over2} (2n+1).
\label{theta1nKKI}
\end{equation}
In particular, $\bar{\theta}_{1,2} = 7.8532\ldots$ 
To understand the relation between 
the solutions $\bar{\theta}_{1,k}$ and the large-$L$ behavior of 
$x_1$ and $x_2$ for 
the low-energy levels of $\hat{C}$, we have performed a detailed 
numerical analysis, computing $x_1(L)$ and $x_2(L)$ as described in 
Sec.~\ref{sec5}. We find $x_{2,n} = 1 + i \bar{\theta}_{1,n}/L$ for the 
$n$-th level of $\hat{C}$, i.e., the $n$-th level correction
term $\theta_{1,n}$ is equal to $\bar{\theta}_{1,n}$. 
In particular, $\bar{\theta}_{1,1}$ and $\bar{\theta}_{1,2}$ refer to the 
ground state and to the first excited state, respectively.
Given that $\theta_{1,n} = \bar{\theta}_{1,n}$, in the following 
we will label the solutions of Eq.~(\ref{eq-cos-cosh}) simply 
with $\theta_{1,n}$.

We can finally compute the gap. Since 
\begin{equation}
\varepsilon(x_1)^2 \approx 4 \gamma^4 + (1-\gamma^2) \theta_1^4 {1\over L^4} 
\label{eps-KKI}
\end{equation}
for large values of $L$, we obtain
\begin{equation}
\Delta_{K/KI} = {1 - \gamma^2 \over 4 \gamma^2} 
(\theta_{1,2}^4 - \theta_{1,1}^4)
   {1\over L^4}.
\label{DeltaKKI}
\end{equation}
While in the K and KI phases the gap decreases as $L^{-2}$, 
at the boundary between the two phases, 
the gap decreases faster, as $L^{-4}$: the dynamic exponent $z$ is equal 
to 4. 

\begin{figure}[tbp]
\begin{center}
\begin{tabular}{c}
\epsfig{width=10truecm,angle=0,file=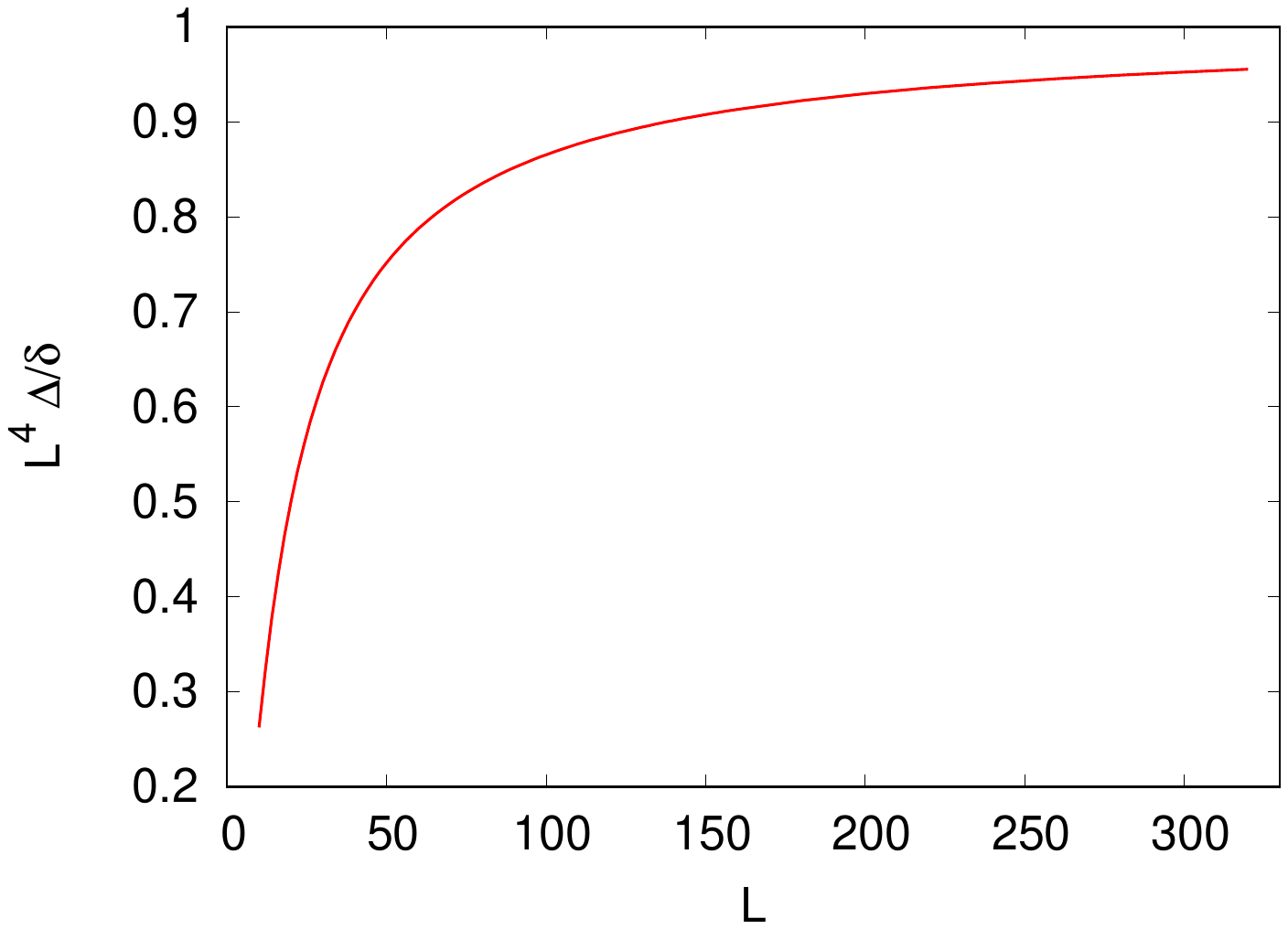}\\
\end{tabular}
\end{center}
\vskip-1cm
\caption{Ratio $\Delta(L)/\Delta_{K/KI}(L)$ as a function
of $L$ for $g = 0.64$, $\gamma = 0.6$, and $J_0 = 0.8$, on the K-KI boundary. 
The gap $\Delta(L)$ has been obtained by a numerical diagonalization of the 
matrix $\hat{C}$, while $\Delta_{K/KI}(L)$ is the asymptotic
expression (\ref{DeltaKKI}).
For these values of the parameters, 
$\Delta_{K/KI}(L) = \delta L^{-4}$ with $\delta = 1467.98$.
}
\label{KKIDelta}
\end{figure}

To verify the predicted behavior, Eq.~(\ref{DeltaKKI}), we have 
considered points on the K-KI boundary and computed numerically 
the gap $\Delta(L)$ for different sizes (we determine it by 
a direct diagonalization of the matrix $\hat{C}$). 
As expected, the ratio $\Delta(L)/\Delta_{K/KI}$
approaches 1 as $L$ increases (numerically we find $1/L$ corrections),
see Fig.~\ref{KKIDelta}. 

A critical behavior with $z=4$ was also observed in an
antiferromagnetic XY model with frustrated boundary conditions
\cite{SCCFSZ-23}. 
It appears to be a generic behavior for $g = g_{K/KI}$ in  
XY models. Indeed,
for this value of $g$ the dispersion relation (\ref{dispersion})
has an expansion $\epsilon(q)^2 = a + b q^4$ for $q\to 0$ (this is 
Eq.~(\ref{eps-KKI}) with $q = \theta_1/L$). The quartic dispersion
relation implies  $z = 4$.

\subsection{Crossover behavior across the K-KI boundary line}
 
Let us now discuss the crossover behavior as the K-KI boundary 
is crossed by varying $g$ at fixed $\gamma$ and $J_0$. We parametrize 
the coupling $g$ in terms of a scaling variable $\epsilon_K$ as 
\begin{equation}
   g = (1 - \gamma^2) \left(1 + {\epsilon_K\over L^\alpha} \right),
 \label{scalingKKI}
\end{equation}
where $\alpha$ is an exponent that we determine below.
We expand $x_1$ and $x_2$ as in Eq.~(\ref{expx1x2KKI}), and substitute 
all expressions in the equation $\varepsilon(x_1)^2 = \varepsilon(x_2)^2$.
We obtain 
\begin{equation} 
  - {4\over L^\alpha} \epsilon_K - {1\over L^2}(\theta_1^2 - x_{11}^2) + 
    O(L^{-3}) = 0. 
\end{equation}
Thus, a nontrivial scaling behavior is obtained by taking 
$\alpha = 2$ and requiring 
\begin{equation}
   x_{11}^2 = 4 \epsilon_K + \theta_1^2.
\label{equationepsilonK}
\end{equation}
Note that, in Eq.~(\ref{expx1x2KKI}), we are not 
assuming that $x_{11}$ is real. This is correct in the K 
phase, but it is not valid deep in the KI phase
where both $x_1$ and $x_2$ are phases.  

An additional relation between $x_{11}$ and $\theta_1$ is obtained as before. 
We consider the set of equations, Eq.~(\ref{eq:foureqs}), set 
$c_2 = 1$ and eliminate $c_1$, $d_1$, and $d_2$. 
After a lengthy calculation, we finally obtain 
the equation
\begin{eqnarray}
 && \cos\theta_1 \cosh x_{11} - 1 = 2 \epsilon_K  
 {\sin \theta_1 \over \theta_1} {\sinh x_{11}\over x_{11}}.
\label{eqx11k0sol} 
\end{eqnarray}
For $\epsilon_K = 0$, we reobtain Eq.~(\ref{eq-cos-cosh}).
This equation, together with Eq.~(\ref{equationepsilonK}), allows
one to compute $x_{11}$ and $\theta_1$ for each value of $\epsilon_K$.
The equations have an 
infinite number of solutions that are directly related with 
the levels of the matrix $\hat{C}$. 
We label the solutions as  $\theta_{1,n}$ and $x_{11,n}$, $n \ge 1$: they are 
ordered so that $0 < \theta_{1,1} < \theta_{1,2} < \ldots$. The values 
$x_{11,n}$ and $\theta_{1,n}$ are obviously
related by Eq.~(\ref{equationepsilonK}). As it is implicit in the 
notation, $\theta_{1,n}$ and $x_{11,n}$ are the values of $\theta_1$ and 
$x_{11}$ for the $n$-th level of the matrix $\hat{C}$. In particular,
the values $x_{11,1}$ and $\theta_{1,1}$ correspond to the 
ground state, while $x_{11,2}$ and $\theta_{1,2}$ 
correspond to the first excited state.

\begin{figure}[tbp]
\begin{center}
\begin{tabular}{c}
\epsfig{width=10truecm,angle=0,file=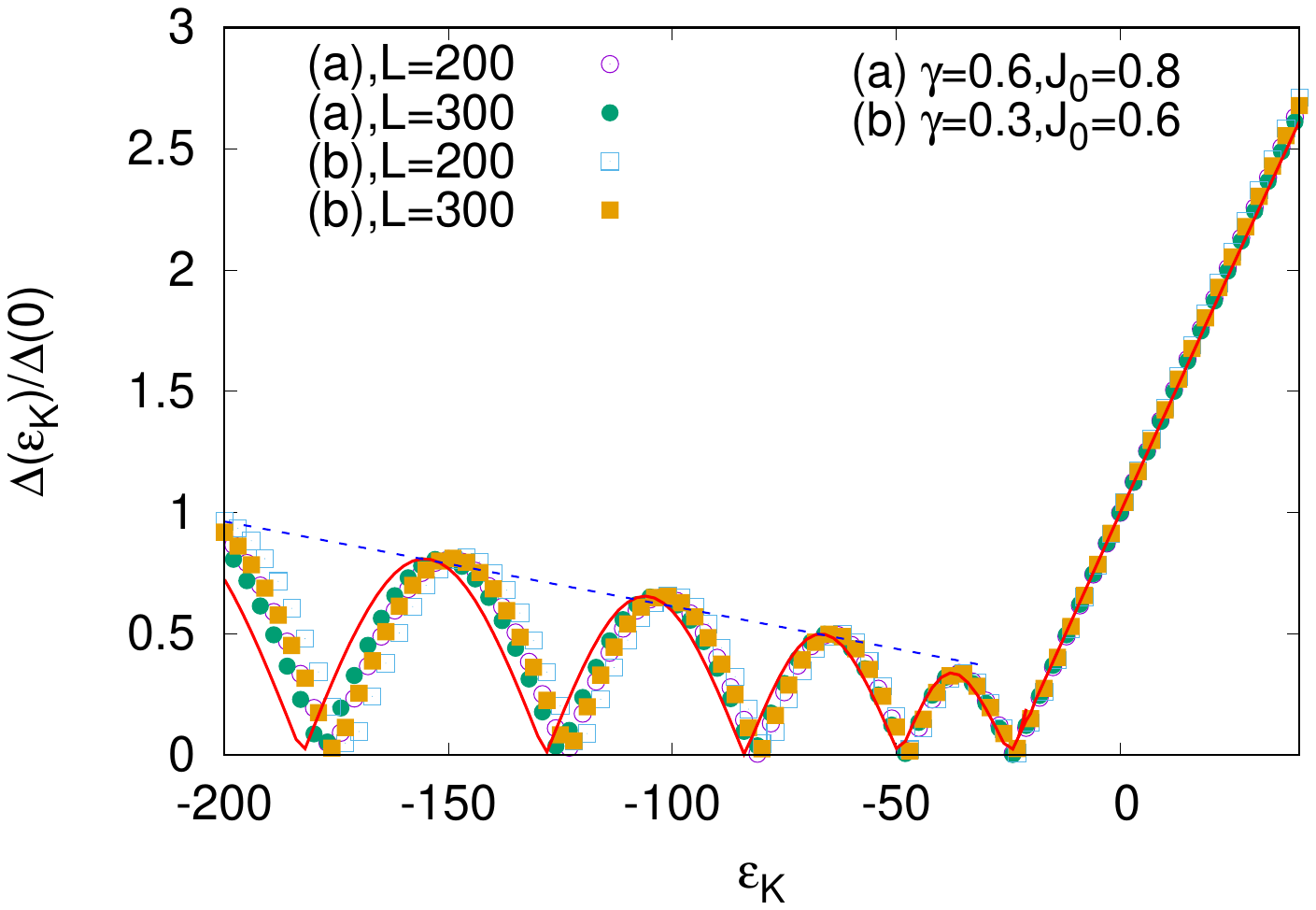}\\
\end{tabular}
\end{center}
\vskip-1cm
\caption{Crossover function across the K-KI boundary. 
We report the finite-size gap ratio $\Delta(\epsilon_K,L)/\Delta(0,L)$ versus 
$\epsilon_K = L^2(g/(1-\gamma^2) - 1)$ for two different 
systems: (a) $\gamma = 0.6$, $J_0 = 0.8$; (b) $\gamma = 0.3$, $J_0 = 0.6$.
The gap $\Delta(\epsilon_K,L)$ 
has been obtained by a numerical diagonalization of the 
matrix $\hat{C}$.
The solid curve is the scaling function $f_{K/KI}(\epsilon_K)$. 
In the KI phase ($\epsilon_K < 0$) the 
crossover function vanishes for $\epsilon_K=-24.674,-49.348,-83.8916,-128.305,
-182.588,\dots$, see Eq.~(\ref{epsiloKfzero}).
The dashed line corresponds to a fit of the maxima of the scaling 
curve for $\epsilon_K < -50$, see text.
}
\label{KKIcrossover}
\end{figure}

Finally, the large-size behavior of the gap is given by 
\begin{equation}
\Delta = {(1-\gamma^2)  \over 2 \gamma^2} (\theta_{1,2}^2 -\theta_{1,1}^2)
   (4 \epsilon_K + \theta_{1,2}^2 +\theta_{1,1}^2)\,  {1\over L^4}.
\end{equation}
As it occurs for $\epsilon_K = 0$, the gap decreases as $L^{-4}$ 
for generic values of $\epsilon_K$. 

The previous results allow us to predict the scaling behavior 
\begin{equation}
\Delta(\epsilon_K,L) = \Delta_0(L) f_{K/KI}(\epsilon_K),
\label{fKKI-def}
\end{equation}
across the K-KI boundary for large values of $L$, 
where $\Delta_0(L)=\Delta_{K/KI}$ is the large-size expression for the gap 
on the boundary line, i.e., for $\epsilon_K = 0$, given in 
Eq.~(\ref{DeltaKKI}). 
In Fig.~\ref{KKIcrossover}, we report the 
scaling curve $f_{K/KI}(\epsilon_K)$ together with numerical results 
for two different systems. Numerical data are in good agreement with the 
asymptotic curve with size corrections that increase as 
$\epsilon_K$ becomes more negative. 
The existence of a universal K-KI scaling allows us to interpret
the K-KI boundary as a surface of continuous transitions.
Correspondingly, in the language of renormalization-group theory, 
$g$, or, more precisely, the difference $g-g_{K/KI}$,
plays the role of relevant perturbation of the  K-KI transition, 
with renormalization-group dimension $y_g = 2$.

Some general properties of the scaling curve $f_{K/KI}(\epsilon_K)$ 
are discussed in \ref{App.KKI}.
For positive $\epsilon_K$ the curve is very well approximated 
by a straight line: $f_{K/KI}(\epsilon_K) \approx a \epsilon_K + b$,
with $a = 0.0356478$, see Fig.~\ref{KKIcrossover}. 
For large negative values of $\epsilon_K$,
$f_{K/KI}(\epsilon_K)$ 
behaves as $\epsilon_K$ times an oscillating bounded function of 
$\epsilon_K$. The data are consistent 
with this behavior: for instance, the maxima of the scaling curve for 
$\epsilon_K < -50$ are well fitted by the linear function
$0.264 - 0.0035 \epsilon_K$, see Fig.~\ref{KKIcrossover}.
Finally, the curve vanishes for 
\begin{equation}
\epsilon_K = -{\pi^2\over 2} (m^2 + 2 m + 2),
\label{epsiloKfzero}
\end{equation}
where $m$ is a positive integer. 

\section{Conclusions} \label{sec10}

In this work we have considered the one-dimensional XY quantum chain in 
a transverse field $g$, see Eq.~(\ref{Isc}), in the presence of 
oppositely-oriented boundary fields (OBF). For $g=1$ 
the model undergoes a continuous transition separating a 
disordered paramagnetic phase ($g> 1$) from 
a low-$g$ phase, in which the bulk phase 
behavior depends on the boundary conditions. 
We determine the energy spectrum of the model. By a generalization
of the approach applied to the Ising chain 
in Ref.~\cite{Hu2021}, the problem of determining the energy spectrum
for a chain of length $L$
is reduced to the problem of solving a system of five algebraic 
equations in five unknowns. The analysis of the solutions of these equations
gives us 
exact analytic results for the energy gap $\Delta$ (the energy difference
between the two lowest energy states) as a function of the 
model parameters for $g < 1$ and large values of $L$. 
On the basis of these results 
we are able to classify the different phases occurring in the low-field
regime. Four different phases emerge: 
\begin{itemize}
\item[i)] A magnetized (M) phase. In this case the system is ferromagnetic,
with two low-lying states. The gap $\Delta$ decreases 
exponentially with the size, as $e^{-a L}$. 
\item[ii)] A magnetized-incommensurate (MI) phase. Also in this case there 
are two low-lying states. The gap behaves as $e^{-a L} f(L)$, where $f(L)$ 
is a bounded oscillating function of $L$. The oscillations are not commensurate
with the system size.
\item[iii)] A kink (K) phase. There is a tower of low-lying delocalized 
states of momenta of order $1/L$, which become degenerate in the 
infinite-volume limit. The gap behaves as $L^{-2}$. 
\item[iv)] A kink-incommensurate (KI) phase. Also in this 
case there is a tower of low-lying delocalized  
excitations that are degenerate in the infinite-volume limit. 
The gap behaves as $L^{-2} f(L)$, where $f(L)$ 
is a bounded oscillating function of $L$. 
\end{itemize}
The M and the MI phases, with two quasidegenerate low-energy states,
are the only ones present when periodic or open boundary conditions 
are considered, see, e.g., Ref.~\cite{Franchini-17}. The K phase 
was already discussed in Ref.~\cite{CPV-15JSTAT} in the context of the 
Ising chain and it appears when the boundary fields are sufficiently strong. 
It can also be induced in Ising rings by sufficiently strong defects 
that destroy the ferromagnetic order \cite{CPV-15}. The KI phase 
is a phase that is only present in the XY model 
for $\gamma < 1$. It has delocalized excitations 
as the K phase, but it also shares with the MI phase the property
that the gap shows incommensurate oscillations with the chain size $L$. 

We have also discussed the behavior of the model along 
the boundaries that separate the different phases. 
Refs.~\cite{CPV-15,CPV-15JSTAT} discussed the crossover behavior 
across the M-K boundary for Ising chains. Here, we perform the 
same calculation for the XY model, confirming the universality of the 
M-K crossover behavior. We have also investigated the behavior along the 
K-KI boundary. On the boundary, we find that the gap decreases as $L^{-4}$,
i.e., that the dynamic critical exponent $z$ is equal to 4. Moreover, 
we are able to completely characterize the crossover across the 
boundary, determining the relevant scaling variable and the 
crossover scaling function for the gap. As expected, the
scaling function is universal, provided the nonuniversal normalization
of the scaling variable is appropriately chosen.

In this paper we have focused on the finite-size behavior of the gap for finite
chains with OBF, but results also apply to XY rings with defects with minor
changes, as discussed in Ref.~\cite{CPV-15}. Moreover, 
it should be possible to extend these results to correlation functions,
using the general techniques of Ref.~\cite{BMD-70,BM-71}, or the perturbative 
approach of Ref.~\cite{CPV-15JSTAT}. 
In particular, it would be interesting to compute the two-point correlation
function and $\langle \sigma_x^{(i)}\rangle$ 
as a function of $x$. Results for the KI phase or, at least for the K-KI boundary, 
would provide a better understanding of  the origin of the size 
oscillations of the gap. One could also determine the entanglement properties
\cite{CCD-09} of the phases, to understand their dependence on the 
boundary interactions \cite{ALS-09}, 
as well as the dynamic behavior under different  
dynamic protocols \cite{RV-21,PV-24}, extending the results of 
Refs.~\cite{PRV-18,PRV-20}. From a technical point of view, our method is quite general
and it can also be applied to the antiferromagnetic XY model in a transverse field
with different types of boundary conditions. In this case we expect 
(for a discussion of these issues, see Refs.~\cite{MGKF-20,TMFG-21,MGF-22}
and references therein)
a rich phase behavior, with different frustrated phases that can be 
stabilized by a proper choice of the boundary conditions and of the 
(even or odd) length of the chain. 
In particular, we should mention the recent results of 
Ref.~\cite{SCCFSZ-23} for the antiferromagnetic XY model
with frustrated boundary conditions. Their model presents two phases
which we can identify with the K and KI phases discussed here and which 
are separated by a transition line with dynamic exponent $z = 4$ as
we observe on the K-KI boundary. It would be interesting to 
perform a more detailed comparison of the two models 
(for instance, one might compare the K-KI crossover function
in the two models) to verify the expected universality.

\bigskip

A.P.  acknowledges support from project PRIN 2022 ``Emerging
gauge theories: critical properties and quantum dynamics''
(20227JZKWP). 

\appendix 

\section{Jordan-Wigner representation and Hamiltonian diagonalization}
\label{App.A}

To compute the spectrum of Hamiltonian (\ref{def-Hextended}),
we follow Ref.~\cite{LSM-61}. We first
perform a Jordan-Wigner transformation, defining fermionic operators
$c_i$ and $c_i^\dagger$, 
\begin{equation}
c_i^\dagger = R_i \sigma_i^+, \qquad c_i = R_i \sigma_i^- ,
\qquad
R_i = (-1)^{i-1}\prod_{j=0}^{i-1} \sigma_j^{(3)},
\label{c-cdag-def}
\end{equation}
where $\sigma^\pm = (\sigma^{(1)} \pm i\sigma^{(2)})/2$. These
relations can be inverted, obtaining
\begin{equation}
\sigma_i^{(1)} = R_i (c_i^\dagger + c_i),  \qquad
\sigma_i^{(2)} = -i R_i (c_i^\dagger - c_i),  \qquad
\sigma_i^{(3)} = 2 c_i^\dagger c_i - 1 .
\label{sigma-c-relation}
\end{equation}
In terms of the fermionic operators, 
Hamiltonian (\ref{def-Hextended}) becomes 
\begin{eqnarray}
H_e &=& - \sum_{i=1}^{L-1} \left[
    c_{i+1}^\dagger c_i + c_i^\dagger c_{i+1} + 
    \gamma (c_i^\dagger c_{i+1}^\dagger - c_i^\dagger c_{i+1})\right] -
    g \sum_{i=1}^{L} (2 c_i^\dagger c_i - 1) \nonumber \\
    && - \sum_{i=0,L} J_i (c_{i+1}^\dagger c_i + c_i^\dagger c_{i+1} + 
     c_i^\dagger c_{i+1}^\dagger + c_{i+1} c_i)
\end{eqnarray}
where the last term is the boundary contribution ($i$ takes the 
values 0 and $L$ only). The previous expression can be rewritten in the 
more compact form
\begin{equation}
H_e = L g - \sum_{i,j=0}^{L+1} \left[c_i^\dagger A_{ij} c_j +
       {1\over2} c_i^\dagger B_{ij} c_j^\dagger +
       {1\over2} c_i B_{ij} c_j \right],
\end{equation}
where the matrices $A$ and $B$ are symmetric and antisymmetric, respectively.
Finally, we perform a Bogoliubov transformation, introducing new
canonical fermionic variables
\begin{equation}
   \eta_k = \sum_{i=0}^{L+1} (g_{ki} c_i^\dagger + h_{ki} c_i),
\end{equation}
where $g_{ki}$ and $h_{ki}$ are fixed by the requirement that $H_e$
takes the form
\begin{equation}
   H_e = E_{gs} + \sum_{k=0}^{L+1} {\cal E}_k \eta^\dagger_k \eta_k,
\label{H1-diag}
\end{equation}
with $0\le {\cal E}_0 \le {\cal E}_1 \le \ldots$
Following Ref.~\cite{LSM-61}, we define the vectors
\begin{eqnarray}
 U_k = (g_{k0} + h_{k0}, g_{k1} + h_{k1},\ldots)  ,\qquad
 V_k = (g_{k0} - h_{k0}, g_{k1} - h_{k1},\ldots)  .
\end{eqnarray}
The variables $\eta_k$ satisfy canonical anticommutation relations if
the vectors $U_k$ form an orthonormal basis, and so does the set
$V_k$.  The vectors $V_k$ satisfy
\begin{equation}
  (A+B)(A-B) V_k = {\cal E}_k^2 V_k.
\end{equation}
Thus, if we define $C = (A+B)(A-B)$, 
the determination of the energies ${\cal E}_k$ is equivalent to the 
determination of the eigenvalues of the matrix $C$. Note that the 
Hamiltonian $H_e$ is invariant under a parity transformation generated
by $P_z = \prod_{i=0}^{L+1} \sigma_i^{(3)}$) and
thus the spectrum is doubly degenerate. This implies that $C$ has 
necessarily a zero eigenvalue. The matrix $C$ is given 
in Eq.~(\ref{eq:defCmatrix}) in the text.

The previous calculations allow us to obtain the spectrum of $H_e$.
We should now discuss how to use these results to obtain the spectrum 
of $H$ in the presence of OBF.
Let us first 
note that the Hamiltonian $H_e$ commutes with both $\sigma_0^{(1)}$ and
$\sigma_{L+1}^{(1)}$, which can therefore be simultaneously
diagonalized.  The Hilbert space can be divided into four sectors, which
we label as $(1,1)$, $(-1,1)$, $(1,-1)$ and $(-1,-1)$, where
$(s_0,s_{L+1})$ are the eigenvalues of $\sigma_0^{(1)}$ and
$\sigma_{L+1}^{(1)}$. The restriction of $H_e$ to each sector gives
rise to the Hamiltonian $H$, defined in Eq.~(\ref{Isc}), with a
boundary term of the form (\ref{hb}). In particular, the restriction
to the sector $(1,-1)$ gives the 
spectrum in the presence of OBF, while the restriction to the sector (1,1)
gives the spectrum with parallel boundary fields.
To conclude the calculation we should determine the states that belong to 
the sector $(1,-1)$. The analysis is reported in Ref.~\cite{CPV-15JSTAT}.
In the sector $s_0 = 1$, $s_{L+1} = -1$, the lowest energy
state is the first excited state $\eta_1^\dagger|0\rangle$ of the
Hamiltonian $H_e$ and all states are obtained as $\eta_{k_1}^\dagger \ldots
\eta_{k_m}^\dagger |0\rangle$ with $k_i \ge 1$ and $m$ odd.
In particular, the first
excited state in the sector is $\eta_2^\dagger|0\rangle$, so that
the energy gap is $\Delta = {\cal E}_2 - {\cal E}_1$.

\section{Consistency of the parametrization of the eigenvectors}
\label{App.consistency}

In Sec.~\ref{sec4} we proved that a vector of the form (\ref{psi-Ansatz})
 is an eigenvector of $\hat{C}$ if the coefficients 
$c_1$, $c_2$, $d_1$, $d_2$ satisfy the 
equations (\ref{eq:foureqs}) and $\varepsilon(x_1)^2 = \varepsilon(x_2)^2$.
In this appendix we wish to prove the opposite result: any eigenvector of 
$\hat{C}$ can be written as in Eq.~(\ref{psi-Ansatz}) with 
appropriate coefficients $c_1$, $c_2$, $d_1$, $d_2$, $x_1$, and $x_2$. 

Let us consider an eigenvector $\phi_L$ of $\hat{C}$ with eigenvalue 
${\cal E}^2$. We wish to show that we can determine 
$c_1$, $c_2$, $d_1$, $d_2$, $x_1$, and $x_2$, so that $\psi_L = \phi_L$. 
We first determine $x_1$ and $x_2$ so that 
$\varepsilon(x_1)^2 = \varepsilon(x_2)^2 = {\cal E}^2$ 
(this is discussed in detail in Sec.~\ref{sec5}). 
Then, we determine $c_1$, $c_2$, $d_1$, $d_2$ by solving the 
linear system of equations $\psi_{L,i} = \phi_{L,i}$ for $i=1,2,3,4$. 

We wish now to prove that the resulting vector $\psi_{L}$ is equal to $\phi_L$, 
i.e., that $\psi_{L,i} = \phi_{L,i}$ for $i \ge 5$. Let us notice that 
Eq.~(\ref{eq:foureqs}) can be reinterpreted as a recursion relation for the 
vector components. For $k \ge 5$, taking into account the particular structure of 
$\hat{C}$, we can write
\begin{equation}
\hbox{Eq}_{k-2} = \sum_{i=k-4}^k C_{k-2,i} \psi_{L,i} - {\cal E}^2 \psi_{L,k-2} = 0, 
\end{equation}
which gives 
\begin{equation}
\psi_{L,k} = {1\over C_{k-2,k}} \left[
   {\cal E}^2 \psi_{L,k-2} - \sum_{i=k-4}^{k-1} C_{k-2,i} \psi_{L,i}\right].
\label{recursionPsi}
\end{equation}
Since the parametrization (\ref{psi-Ansatz}) satisfies $\hbox{Eq}_{k} = 0$
for $3\le k \le L-1$ (therefore, $\hbox{Eq}_{k-2} = 0$ holds for 
$5\le k \le L+1$), we can use Eq.~(\ref{recursionPsi}) to determine all 
components $\psi_{L,k}$, $k\ge 5$, in terms of $\psi_{L,i}$ with $i=1,2,3,4$. 
Recursion (\ref{recursionPsi}) (with $\phi$ replacing $\psi$) 
also holds for $\phi_L$ 
(it satisfies $\hbox{Eq}_{k} = 0$ for any $k$) and thus the same recursion 
gives $\phi_{L,k}$ in terms of $\phi_{L,i}$ with $i=1,2,3,4$. 
It is then enough to note that the starting values of the recursion are the 
same ($\psi_{L,i} = \phi_{L,i}$ for $i=1,ldots, 4$)
to conclude that $\phi_{L,k} = \psi_{L,k}$ for any $k$.

\section{Magnetized phases: Determination of the parameters $x_i$}
\label{App.magnetized}

In this Appendix we wish to determine the values that $x_1$ and $x_2$ take 
for $L\to \infty$, when the system is in a magnetized phase. 
Here $x_1$ and $x_2$ always refer to the (degenerate) ground state.

As discussed in Sec.~\ref{sec5},
in the magnetized phases $|x_1|$ and $|x_2|$ are both larger than $1$. 
Therefore, for large system sizes, the quantities
$x_1^{-L}$ and $x_2^{-L}$ are exponentially small and thus the leading 
behavior is obtained by neglecting these terms. \footnote{More precisely,
these terms can be neglected provided that the coefficients $c_i$ and $d_i$ 
do not increase exponentially with the system size. We have verified 
this assumption numerically: we compute the eigenvectors by diagonalizing
the matrix $\hat{C}$ and then determine the coefficients $c_1$, $c_2$, $d_1$
and $d_2$ as explained in \ref{App.consistency}. The analytic results 
presented in this Appendix are consistent with this assumption.}
Equations (\ref{eq:foureqs}) decouple: 
$\hbox{Eq}_1$ and $\hbox{Eq}_2$ depend only on $c_1$ and $c_2$, 
while $\hbox{Eq}_L$ and $\hbox{Eq}_{L+1}$ depend only $d_1$ and $d_2$. 

There are two classes of solutions of $\hbox{Eq}_L=0$ and $\hbox{Eq}_{L+1}=0$.
First, there is the trivial solution $d_1 = d_2 = 0$. If this is not the 
case,  we obtain 
\begin{equation}
 d_2 = - \left({x_1\over x_2}\right)^3 d_1\; 
\label{magn_eq1_app}
\end{equation}
and the relation
\begin{equation}
  x_1 x_2 = {(1 + \gamma)^2 - 4 J_0^2 \over 1 - \gamma^2}  \; .
\label{magn_eq2_app}
\end{equation}
Equations $\hbox{Eq}_1=0$ and $\hbox{Eq}_2=0$ have a larger set 
of solutions. Beside the trivial solution $c_1 = c_2 = 0$, we have 
four different possibilities:
\begin{itemize}
\item $c_1 = 0$ and $f_1(x_2) = 0$, with $f_1(x_1) \not =0$; 
\item $c_2 = 0$ and $f_1(x_1) = 0$, with $f_1(x_2) \not =0$; 
\item $f_1(x_1) = 0$ and $f_1(x_2) =0$;
\item $x_1$ and $x_2$ satisfy Eq.~(\ref{magn_eq2_app}) and 
\begin{equation}
c_2 = - c_1 {x_1[1 - \gamma - 2 g x_1 + (1 + \gamma) x_1^2] 
    \over 
    x_2[1 - \gamma - 2 g x_2 + (1 + \gamma) x_2^2]}.
\label{magn_eq4}
\end{equation}
\end{itemize}
Here, the function $f_1(x)$ is defined as
\begin{equation}
f_1(x) = (1 + \gamma) x^2 - 2 g x + (1 - \gamma) = 0.
\end{equation}
Finally, we should determine the additional constraints due to the 
fact that $x_1$ and $x_2$ should satisfy the equation
$\varepsilon(x_1)^2 = \varepsilon(x_2)^2$, which  we rewrite in the  
more convenient form 
\begin{equation}
  x_1^2 x_2^2 [\varepsilon(x_1)^2 - \varepsilon(x_2)^2] = 0.
\label{eqB5}
\end{equation}
Let us first assume that $x_1$ and $x_2$ satisfy Eq.~(\ref{magn_eq2_app}).
We rewrite each term $x_1^a x_2^b$ as  $(x_1 x_2)^K x_1^{a-K} x_2^{b-K}$,
with $K = \min(a,b)$ and then we use Eq.~(\ref{magn_eq2_app}) for 
$x_1 x_2$. Eq.~(\ref{eqB5}) drastically simplifies, allowing us to 
obtain 
\begin{equation}
  x_1 + x_2 = {2 g [(1 + \gamma)^2 - 4 J_0^2] \over 
      (1 - \gamma^2) (1 + \gamma - 2 J_0^2)}.
\label{magn_eq3_app}
\end{equation}
A second possibility is that $x_1$ satisfies $f_1(x_1) = 0$.
If this is the case, we obtain that 
$x_2$ satisfies either $f_1(x_2) = 0$ or $f_1(1/x_2) = 0$. 
Analogously, if $f_1(x_2) = 0$, $x_1$ satisfies either $f_1(x_1) = 0$ or 
$f_1(1/x_1) = 0$.  Combining all results we conclude that there are two
possibilities for 
$x_1$ and $x_2$ in the magnetized phase. They should satisfy
one of these two conditions:
\begin{itemize}
\item Eq.~(\ref{magn_eq2_app}) and (\ref{magn_eq3_app});
\item $f_1(x_i) f_1(1/x_i)= 0$ for both $x_1$ and $x_2$.
\end{itemize}
We have not been able to understand analytically
which of the two conditions is the 
relevant one for each value of the parameters $g$, $\gamma$, and 
$J_0$ and we have thus performed a numerical analysis
as discussed in Sec.~\ref{sec5}. We determine numerically the eigenvalues 
of the matrix $\hat{C}$, and then obtain the values $x_1$ and $x_2$ for the 
ground state by solving $\varepsilon(x)^2 = {\cal E}_1^2$. Finally, we check 
which of the two conditions is satisfied.
The result is particularly simple. For $\gamma > 0$, 
$x_1$ and $x_2$ satisfy Eqs.~(\ref{magn_eq2_app}) and (\ref{magn_eq3_app}). 
On the other hand, for $\gamma < 0$, $x_1$ and $x_2$ both satisfy 
$f_1(x_i) f_1(1/x_i) = 0$. 

The previous equations allow us to determine $x_1$ and $x_2$, as long
as they both satisfy $|x_i|>1$.  If the solutions are both real and 
larger than 1 in absolute value, the system is in 
M phase for the given set of parameters. If they are complex conjugate 
with $|x_1| = |x_2| > 1$, the system is in the MI phase. 
    
It is important to note that
we obtain a two-dimensional space of solutions. 
Indeed, if $x_1$ and $x_2$ are computed by using Eqs.~(\ref{magn_eq2_app}) and 
(\ref{magn_eq3_app}), we can arbitrarily fix $c_1$ and $d_1$ and compute
$c_2$ and $d_2$ using Eqs.~(\ref{magn_eq1_app}) and 
(\ref{magn_eq4}). On the other hand, if $f_1(x_i)f_1(1/x_i) = 0$, 
we should set $d_1 = d_2 = 0$ and arbitrarily choose $c_1$ and 
$c_2$, or, viceversa, set $c_1 = c_2 = 0$ and choose $d_1$ and $d_2$. 
Therefore, if we neglect  exponentially small terms,
the solutions form a degenerate two-dimensional space.
This is exactly the expected behavior 
for the two lowest-energy states in a magnetized phase.  The degeneracy is 
lifted for finite values of $L$. To compute the splitting, one should 
consider the terms of order $x_1^{-L}$ and $x_2^{-L}$. Since 
$|x_2| < |x_1|$, the dominant contribution is due to $x_2$. 
We thus predict the gap to scale as $|x_2|^{-L}$, with incommensurate 
oscillations if $x_2$ is complex.

\section{The kink phase} \label{App.kink}

To determine the behavior of the gap in the kink phase 
in which $x_1$ is real and 
larger than 1 in absolute value, and $x_2=e^{i \phi}$,
we have first performed a numerical study of the size behavior of 
the phase $\phi$.  We diagonalize $\hat{C}$ and then 
compute $x_{1,n}(L)$ and $x_{2,n}(L)$ for the 
low-lying energy states (the ``kink" states) 
by solving the equation $\varepsilon(x)^2 = {\cal E}_n$. 
We find that for all these states 
$\phi$ vanishes in 
the infinite-size limit as $1/L$, i.e., 
$\phi_n \approx \theta_{1.n}/L$ for large
sizes (as usual, the suffix $n$ indicates that 
$\phi_n$ is the values of $\phi$ for the $n$-th level of the 
matrix $\hat{C}$). 
Moreover, inside the kink phase $\theta_{1,n}$ never vanishes.

On the basis of these numerical results, 
we parametrize $\phi$ for the lowest-lying states as 
\begin{equation}
  \phi = {\theta_1\over L} + {\theta_2\over L^2} + O(L^{-3}). 
  \label{expansion-phi-app}
\end{equation}
This specific behavior of $x_2$ allows us to determine $x_1$ in the 
infinite-size limit from the 
equation $\varepsilon(x_1)^2 = \varepsilon(x_2)^2$. Setting $x_2 = 1$,
we obtain 
\begin{equation}
   x_1 = x_{10} \equiv {1\over 1 - \gamma^2} 
   [2 \sqrt{g (\gamma^2 + g - 1}) + \gamma^2 + 2 g - 1],
\label{kink-eq1-app}
\end{equation}
a result that is fully supported by the numerical results obtained 
by the direct diagonalization of $\hat{C}$.  Corrections
decay as $1/L^2$ as described below. 

To determine the correction term $\theta_1$,  we expand $x_1$ as 
\begin{equation}
   x_1 \approx  x_{10} + {x_{11}\over L} + {x_{12}\over L^2}.
\end{equation}
Substituting this expansion and Eq.~(\ref{expansion-phi-app}) in 
the equation $\varepsilon(x_1)^2 = \varepsilon(x_2)^2$, we obtain 
$x_{11} = 0$ at order $1/L$, and $x_{12} = A \theta_1^2$ 
($A$ is a function of $g$ and $\gamma$) at order $1/L^2$. Thus, at order 
$1/L$ we can approximate $x_1$ with the leading term $x_{10}$. 
Then, we consider the four equations $\hbox{Eq}_k=0$ reported in 
Eq.~(\ref{eq:foureqs}).  We drop the terms proportional
to $x_1^{-L}$, which are exponentially small (in the kink phase $|x_1|>1$)
and replace $x_1$ with $x_{10}$. The coefficients $c_1$, $c_2$, $d_1$, and 
$d_2$ can be determined up to a multiplicative constant (the normalization of 
$\psi$).  We have verified numerically that $c_2$
never vanishes for generic values of the parameters in the kink phase,
and therefore we can use the above freedom to set $c_2 = 1$.
Then, we solve the linear equations $\hbox{Eq}_2=0$, 
$\hbox{Eq}_L=0$, and $\hbox{Eq}_{L+1}=0$ in terms of 
$c_1$, $d_1$ and $d_2$ and substitute the results in $\hbox{Eq}_1=0$. 
The equation depends on $x_2$ and $x_2^{-L}$. If we now take the limit 
$L\to \infty$ and use the expansion of $x_2$, we obtain the condition 
\begin{equation}
   A(g,\gamma,J_0) \sin\theta_1 = 0,
\end{equation}
where $A(g,\gamma,J_0)$ is a nontrivial function of the parameters.
It implies 
\begin{equation}
 \theta_1 = k \pi \qquad \phi = {k \pi\over L} + O(L^{-2}).
\label{theta1-app}
\end{equation}
It is not restrictive to assume that $\phi\ge 0$, so that 
$k$ is a nonnegative integer. We will now argue that, for generic points in
the kink phase, $k$ should be strictly positive.  More precisely, 
for the $n$-th level we have $\theta_{1,n} = n \pi$. 
This result is the same as that obtained in the kink phase 
of the Ising chain with $\gamma = 1$ \cite{CPV-15JSTAT}.

To show that $\theta_1$ cannot be zero, we combine
numerical and analytic results. Let us suppose 
the opposite, assuming that
$\phi \approx \theta_k/L^k$ for large $L$, with $k\ge 2$. Equation 
$\hbox{Eq}_1=0$ gives 
\begin{equation}
   B(g,\gamma,J_0) \theta_n L^{1-n} + O(L^{-n}) = 0,
\end{equation}
implying $\theta_k = 0$. Thus, if $\theta_1 = 0$, 
$\phi$ should decrease faster than any power of $L$ (most probably 
exponentially), which in turn would 
imply $x_{1} = x_{10}$, $x_2 = 1$ with corrections that decrease 
faster than any power of $L$. As already mentioned, the numerical analysis 
reported at the beginning of this appendix
allow us to exclude this type of behavior 
inside the kink phase. Thus,  in the kink phase we should 
only consider positive integer values of $k$, as it was proved 
rigorously for the Ising chain \cite{CPV-15JSTAT}.
We have thus fully characterized the values of $x_1$ and $x_2$ that 
correspond to the eigenvectors of $\hat{C}$. 

Relations (\ref{kink-eq1-app}) and (\ref{theta1-app}) also hold on
the boundary between the M phase and the K phase, see Eq.~(\ref{gKM}).
However, numerically we find that $x_2=1$ with essentially no corrections
for the lowest energy state (eigenvalue ${\cal E}_1^2$), implying 
$\theta_{1,1} = 0$. Higher-energy eigenstates  correspond to 
$\theta_{1,n} = (n-1) \pi$ with $n > 1$. This different behavior for points
inside the K phase and on the boundary K-M is not surprising, as 
the same occurs in the Ising chain (for this model an analytic 
proof is given in Ref.~\cite{CPV-15JSTAT}).

\section{The magnetized-kink boundary} 
\label{App.MK}

In this Appendix we wish to show the validity of Eq,~(\ref{Deltascaling-MK})
along the whole M-K boundary  and, in particular, we wish to compute the 
constant $R$ so that the scaling curve $f_\Delta(\zeta_s)$ is the same as 
that computed in Ref.~\cite{CPV-15JSTAT}. This result proves the 
universality of the M-K crossover.

To determine the gap scaling function we should determine the 
relation between $x_1$,$x_2$ and the scaling variable $\zeta_s$. For this
purpose we proceed as in \ref{App.kink}.  We expand $x_1$  and $x_2$
as 
\begin{equation}
x_1 = x_{1,K/M} + {x_{11}\over L} + O(L^{-2}), \qquad
x_2 = 1 + {i \theta_1\over L} + O(L^{-2})
\end{equation}
and set $J_0 = J_{0c} + \epsilon/L$. Note that we have included a factor $i$
in the expansion of $x_2$, to make the expansion look similar to the 
one performed in Ref.~\cite{CPV-15JSTAT} ($\theta_1$ corresponds to $k$ in
Ref.~\cite{CPV-15JSTAT}). However, we are not assuming $\theta_1$ to be real: 
$\theta_1$ will be purely imaginary in the magnetized phase.

As in~\ref{App.kink}, we first verify that $x_{11} = 0$, so that, at order
$1/L$, we can replace $x_1$ with $x_{1,K/M}$ in all equations
(\ref{eq:foureqs}).  Then, we 
set $c_2 = 1$, determine $c_1$, $d_1$ and $d_2$ using three of the 
four equations (\ref{eq:foureqs}), and substitute the 
result in the equation $\hbox{Eq}_1 = 0$. The calculation is quite involved.
At the end we obtain an equation of the form 
\begin{equation}
  A \epsilon \theta_1 \cos\theta_1 + B \theta_1^2 \sin \theta_1 + 
  C \epsilon^2 \sin\theta_1 = 0,
\label{eqKMdiretta}
\end{equation}
where $A$, $B$, $C$ are complicated 
functions of the model parameters. This equation
relates $\theta_1$ with the parameter $\epsilon$. 

We would now like to show that,
by a proper choice of the constant $R$ defined in Eq.~(\ref{defzetas}),
we can reexpress this relation as in Ref.~\cite{CPV-15JSTAT}:
\begin{equation}
  4 \zeta_s \theta_1 + (4 \zeta_s^2 - \theta_1^2) \tan \theta_1 = 0.
\label{eqzetas}
\end{equation}
This would prove that the scaling functions for the XY model are the 
same as those computed for the Ising chain. 

Eqs.~(\ref{eqKMdiretta}) and (\ref{eqzetas}) look similar. 
By comparing the two expressions,  we find that it is possible 
to rewrite Eq.~(\ref{eqKMdiretta}) as Eq.~(\ref{eqzetas}) by a proper
choice of $R$, only if $A$, $B$, and $C$ satisfy the relation
\begin{equation}
   A = - 2 \sqrt{-BC}.
\label{relationABC}
\end{equation} 
We have not been able to verify relation (\ref{relationABC}) analytically.
We have therefore performed a numerical check. We have considered 100 different
boundary points, finding that Eq.~(\ref{relationABC}) is 
verified to machine precision in all cases.
If relation (\ref{relationABC}) holds, the crossover function
is universal, provided the constant $R$ is 
\begin{equation}
   R = {1\over 2} \sqrt{- {C\over B}}.
\label{def-costanteR}
\end{equation}
We have not been able to simplify analytically the expression 
(\ref{def-costanteR}). We will now derive $R$ differently, assuming 
that universality holds and using the results of Ref.~\cite{CPV-15JSTAT}. 
This quick and simple method provides a significantly simpler 
expression for the constant.  For $\zeta_s \to - \infty$, i.e., 
deep in the magnetized phase, the scaling function can be 
expanded as (see Ref.~\cite{CPV-15JSTAT})
\begin{equation}
  f_\Delta(\zeta_s) = {32 \over \pi^2} \zeta_s^2 e^{2 \zeta_s} 
         = {32 \over \pi^2} R^2 L^2 (J_0 - J_{0c})^2 e^{2 R L (J_0 - J_{0c})}.
\end{equation}
In the M phase we expect $\Delta(J) = a x_2^{-L}$ ($a$ depends on all model
parameters), while $\Delta(J_{0c}) = b/L^2$, so that 
\begin{equation}
  f_\Delta(\zeta_s) = {a\over b} L^2 e^{-L \ln x_2}.
\end{equation}
Comparing the two expressions we obtain 
\begin{equation}
   \ln x_2 = - 2 R (J_0 - J_{0c}) 
\end{equation}
in the limit $J_0 \to J_{0c}$. Expanding $x_2$ [obtained by solving 
Eqs.~(\ref{magn_eq2}) and (\ref{magn_eq3})] close to the point 
$J_0 = J_{0c}$, this relation  gives Eq.~(\ref{Rconst}).

\section{Properties of the K-KI crossover function} \label{App.KKI}

The scaling curve $f_{K/KI}(\epsilon_K)$ defined in 
Eq.~(\ref{fKKI-def}) shows two different types of behaviors. For positive 
$\epsilon_K$ the curve is essentially a straight line. This linear behavior 
can be predicted by noting that $\Delta(\epsilon_K)$, computed for 
$\epsilon_K\to \infty$  (deep in the K phase), 
should reproduce Eq.~(\ref{DeltaK}) computed in
the limit $g\to 1 - \gamma^2$ (close to the boundary). This implies 
$f_{K/KI}(\epsilon_K) \approx a \epsilon_K$ for $\epsilon_K\to \infty$,
with
\begin{equation}
 a = {12 \pi^2 \over \theta_{1,2}^4 - \theta_{1,1}^4} \approx 0.0356478.
\end{equation}
where $\theta_{1,2}$ and $\theta_{1,1}$ are the phases on the K-KI boundary,
i.e., for $\epsilon_K = 0$, see Eqs.~(\ref{theta11KKI}) and 
(\ref{theta1nKKI}).\footnote{As before,
$x_{11,n}$ and $\theta_{1,n}$ indicate the values of $x_{11}$ and of $\theta_1$
for the $n$-th level of the matrix $\hat{C}$.}

The approximately linear  behavior is 
observed as long as $x_{11,1}$ and $x_{11,2}$ are both real and positive.
When one or both become complex, the behavior changes. To determine 
the values $\epsilon_{K,n}$ of $\epsilon_K$ where $x_{11,n} = 0$, we first note
that the values $x_{11,n}$ are all
positive for $\epsilon_K>0$. If we decrease $\epsilon_K$, the parameters 
$x_{11,n}$ decrease, as expected on the basis of Eq.~(\ref{equationepsilonK}).
So there is a 
value $\epsilon_K$ (a different one for each $n$, that we indicate with 
$\epsilon_{K,n}$), 
for which $x_{11,n} = 0$. If 
$\epsilon_K$ is further decreased, $x_{11,n}$ becomes complex. 
To determine the values of 
$\epsilon_K$ for which there is a solution with $x_{11} = 0$, 
we set $x_{11}=0$ in Eq.~(\ref{equationepsilonK}), 
which gives $\epsilon_K = - \theta_1^2/4$. Then 
Eq.~(\ref{eqx11k0sol}) becomes
\begin{equation}
1 - \cos\theta_1 = {\theta_1\over 2} \sin \theta_1.
\label{eqE2}
\end{equation}
There are two classes of solutions of this equation: 
(i) $\theta_1 = 2 k \pi$; (ii) the 
solutions of $\tan \theta_1/2 = \theta_1/2$. To understand the relation 
between these solutions and the values $\theta_{1,n}$, we have 
performed a numerical analysis, finding that each solution 
of Eq.~(\ref{eqE2}) corresponds to 
a different level of the matrix $\hat{C}$.
For $\theta_1 = 2 \pi$, we find that the lowest-energy $x_{11,1}$ vanishes.
Thus, $\theta_{1,1} = 2\pi$ and $\epsilon_{K,1} = -\pi^2 = -9.8696$.
The second smallest $\theta_1$ that is a solution
of Eq.~(\ref{eqE2}) is $\theta_1 = 8.98682$ [the smallest 
$\theta_1$ that is a solution of type (ii)].
It corresponds to $x_{11,2} = 0$, so that $\epsilon_{K,2} = -20.1907$.
We thus conclude that all $x_{11,n}$ are positive and real for 
$\epsilon_K > \epsilon_{K,1} = -9.8696$, while, for 
$-20.1907 < \epsilon_K < -9.8696$, $x_{11,1}$ is complex
and $x_{11,2}$ (and also $x_{11,n}$, $n > 2$) is real.
For $\epsilon_K< -20.1907$, 
$x_{11}$ is purely imaginary both for the ground state and for the first 
excited state. In this regime, the scaling function shows an expected 
oscillatory behavior.  

A second interesting property of the scaling function is that 
it vanishes on a sequence of values of 
$\epsilon_K$, where $\theta_{1,1} = \theta_{1,2}$, i.e., the 
correction term is the same for the ground state and the first
excited state. To determine these points, let us rewrite 
Eq.~(\ref{eqx11k0sol}) as $f(\theta_1) = 0$, replacing 
$x_{11}$ with the expression obtained using Eq.~(\ref{equationepsilonK}).
The solutions such that $\theta_{1,1} = \theta_{1,2}$, should 
also satisfy $f'(\theta_1) = 0$, a condition that can be written as 
\begin{eqnarray}
&& (\theta_1^2 + x_{11}^2) (\cos\theta_1 \cosh x_{11} - 1) 
\label{eqder} \\
&& \qquad + x_{11} (\theta_1^2 - 2 \epsilon_K) \cos\theta_1 \sinh x_{11} - 
      \theta_1(x_{11}^2 + 2 \epsilon_K) \sin\theta_1 \cosh x_{11} = 0.
\nonumber 
\end{eqnarray}
Thus, the values of $\epsilon_K$ where the phases of two 
different levels coincide, i.e., $\theta_{1,n}=\theta_{1,n+1}$ for some
$n$, can be determined by solving 
Eqs.~(\ref{equationepsilonK}), (\ref{eqx11k0sol}), and (\ref{eqder}) 
in terms of $\theta_1$, $x_{11}$ and $\epsilon_K$. By inspection, it 
is immediate to verify that $\theta_1 = 2 \pi m_1$, $x_{11} = 2 \pi m_2 i$,
$\epsilon_K = - \pi^2 (m_1^2 + m_2^2)$ is a solution for any integer 
$m_1$ and $m_2$. Numerically, we have verified that all solutions 
that correspond to $\theta_{1,1} = \theta_{1,2}$ belong to this class. 
More precisely, they are a subset with $m_1 = m_2 + 2$.
Therefore, the points with vanishing $L^{-4}$ correction correspond to
\begin{equation}
\epsilon_K = -{\pi^2\over 2} (m^2 + 2 m + 2),
\label{epsiloKfzero-app}
\end{equation}
where $m$ is a positive integer. For these 
values of $\epsilon_K$, $\theta_{1,1} = \theta_{1,2} = 2 \pi (m+2)$ and 
$x_{1,1} = x_{1,2} = 2 \pi m i$.

Finally, let us discuss the behavior of the scaling function in the 
limit $\epsilon_K\to-\infty$, i.e., deep in the KI phase. 
To obtain analytic results, we should use exact results for the gap in the KI 
phase. Unfortunately, we have not been able to determine the exact asymptotic 
behavior of the gap in this limit. Numerically, we find that the 
gap decreases as $L^{-2} f(L)$,
where $f(L)$ is an oscillatory function,
see, e.g., the upper left panel in Fig.~\ref{RescaledGap}. 
If this is correct, we conclude that $f_{K/KI}(\epsilon_K)$ 
behaves as $\epsilon_K$ times an oscillating bounded function of 
$\epsilon_K$ for $\epsilon_K\to-\infty$. The data are consistent 
with this behavior: for instance, the maxima of the scaling curve for 
$\epsilon_K < -50$ are well fitted by the linear function
$0.264 - 0.0035 \epsilon_K$, see Fig.~\ref{KKIcrossover} in the text.

\end{document}